\pdfoutput=1

\documentclass[onecolumn]{emulateapj}

\usepackage{amsmath}
\usepackage{graphicx}
\usepackage{color}
\usepackage[colorlinks]{hyperref}
\hypersetup{
    colorlinks,	
    citecolor=blue,
}

\newcommand{\be}{\begin{eqnarray}}
\newcommand{\ee}{\end{eqnarray}}

\def\jcap{JCAP}
\def\actaa{Acta Astron.}

\shorttitle{Impact of the returning radiation}
\shortauthors{Riaz et al.}

\begin{document}

\title{Impact of the returning radiation on the analysis of the reflection spectra of black holes}

\author{Shafqat~Riaz\altaffilmark{1}, {Micha{\l} Szanecki}\altaffilmark{2}, {Andrzej Nied{\'z}wiecki}\altaffilmark{3}, Dimitry~Ayzenberg\altaffilmark{1,4} and Cosimo~Bambi\altaffilmark{1,\dag}, }

\altaffiltext{1}{Center for Field Theory and Particle Physics and Department of Physics, Fudan University, 200438 Shanghai, China. \email[\dag E-mail: ]{bambi@fudan.edu.cn}}
\altaffiltext{2}{Nicolaus Copernicus Astronomical Center, Polish Academy of Sciences, PL-00-716 Warszawa, Poland}
\altaffiltext{3}{Faculty of Physics and Applied Informatics, {\L}{\'o}d{\'z} University, PL-90-236 {\L}{\'o}d{\'z}, Poland}
\altaffiltext{4}{Theoretical Astrophysics, Eberhard-Karls Universit\"at T\"ubingen, 72076 T\"ubingen, Germany}

\begin{abstract}
A fraction of the electromagnetic radiation emitted from the surface of a geometrically thin and optically thick accretion disk of a black hole returns to the disk because of the strong light bending in the vicinity of the compact object (returning radiation). While such radiation clearly affects the observed spectrum of the source, it is often neglected in theoretical models. In the present paper, we study the impact of the returning radiation on relativistic reflection spectra. Assuming neutral material in the disk, we estimate the systematic uncertainties on the measurement of the properties of the system when we fit the data with a theoretical model that neglects the returning radiation. Our \textsl{NICER} simulations show that the inclination angle of the disk and the black hole spin parameter tend to be overestimated for low viewing angles, while no clear bias is observed for high viewing angles. The iron abundance of the disk is never overestimated. In the most extreme cases (in particular, for maximally rotating black holes) the returning radiation flattens the radial emissivity beyond a few gravitational radii. In such cases, it also produces residuals that cannot be compensated by adjusting the parameters of models neglecting the returning radiation. This may be an important issue for interpretation of data from future X-ray missions (e.g. \textsl{Athena}). When we simulate some observations with \textsl{NuSTAR} and we fit data above 10~keV, we find that some conclusions valid for the \textsl{NICER} simulations are not true any longer (e.g., we can get a high iron abundance).
\end{abstract}


%%%%%%%%%%%%%%%%%%%%%%%%%%%%%%%

\section{Introduction}

The X-ray spectra of accreting black holes are often characterized by a power law component, generated by inverse Compton scattering of thermal photons from the accretion disk off free electrons in a hot corona~\citep{1979Natur.279..506S}, and a blurred reflection component, produced by the illumination of the disk by the Comptonized photons~\citep{1991MNRAS.249..352G,2005MNRAS.358..211R,2010ApJ...718..695G}. The reflection spectrum of the disk can be extremely informative about the accretion process in the strong gravity region, and can be analyzed to measure the properties of the system~\citep{2006ApJ...652.1028B,2013Natur.494..449R,2013ApJ...775L..45M,2014ApJ...780...78T,2014ApJ...787...83M,2019SCPMA..6229504D} as well as to test Einstein's theory of general relativity in the strong field regime~\citep{2003IJMPD..12...63L,2009GReGr..41.1795S,2013ApJ...773...57J,2017RvMP...89b5001B,2017ApJ...842...76B,2018PhRvL.120e1101C,2019ApJ...875...56T,2019ApJ...884..147Z}.

Precision measurements of accreting black holes using X-ray reflection spectroscopy require high quality data and sufficiently sophisticated theoretical models. While there has recently been remarkable progress in the development of relativistic reflection models, all the available models still have a number of simplifications that introduce systematic uncertainties in the final measurements and thus limit our capability of precisely measuring the properties of accreting black holes. Generally speaking, modeling uncertainties can arise from: $i)$ the calculation of the reflection spectrum in the rest-frame of the gas in the disk, $ii)$ the description of the accretion disk, $iii)$ the description of the hot corona, and $iv)$ relativistic effects not taken into account. Depending on the specific properties of the source, the spectral state of the source during the observation, the quality of the data, and which of the physical quantities of the system we want to measure, some simplifications in the theoretical model may be justified in some circumstances and introduce unacceptably large systematic errors in other cases. In order to have the systematic uncertainties under control, it is crucial to have at least a rough estimate of the impact of every modeling simplification in the final measurement of the physical properties of the system.

Among the simplifications listed above, the impact of the disk structure and of the coronal geometry are normally thought to be the most important and have received somewhat more attention. In the available relativistic reflection models, the accretion disk is supposed to be geometrically thin and optically thick, and it is often approximated as infinitesimally thin. \citet{2018ApJ...855..120T} and \citet{2020arXiv200309663A} constructed relativistic reflection models for thin disks of finite thickness and studied the impact of the disk thickness on the measurement of the properties of a source. \citet{2020MNRAS.491..417R,2019arXiv191106605R} investigated the modeling bias when we employ an infinitesimally thin disk model to fit data of a source with a thick accretion disk. In all these cases, a crucial ingredient is the coronal geometry and, in turn, the illumination of the accretion disk. Different coronal geometries lead to different disk emissivity profiles~\citep{2003MNRAS.344L..22M,WF2012,2015MNRAS.449..129W}, 
and the latter may cause either negligible or unacceptably large modeling bias in the final measurements~\citep{2018ApJ...855..120T,2020arXiv200309663A,2020MNRAS.491..417R,2019arXiv191106605R}.

\citet{2008ApJ...675.1048R} simulated thin accretion disks in a pseudo-Newtonian potential to study the impact of the thickness of the disk and of the reflection radiation emitted from the plunging region on black hole spin measurements, finding that black hole spins tend to be overestimated, but that the systematic error decreases as the black hole spin increases. \citet{2020arXiv200506719C} estimated the impact of the radiation from the plunging region on the capability of testing the Kerr metric, finding that for small plunging regions the effect is negligible, while larger and larger systematic uncertainties are present as the size of the plunging region increases. The presence of magnetic fields may also have an impact on the accretion disk structure~\citep{2010ApJ...711..959N,2012MNRAS.420..684P} and, in turn, on the reflection spectrum from the disk~\citep{2014JCAP...07..059F}.

Concerning point $iv)$, relativistic effects not taken into account, they are normally thought not to be particularly large. For example, if the plunging region were optically thin, the reflection component of the source would include the reflection radiation produced by the other side of the disk and the reflection radiation circling the black hole one or more times due to the strong light bending in the vicinity of the compact object. \citet{2020PhRvD.101d3010Z} studied the impact of such a radiation, concluding that the effect can be safely ignored in theoretical models for the analysis of present and near-future observational data. Moreover, \citet{1997ApJ...488..109R} and \citet{2020MNRAS.493.5532W} find that the plunging region remains optically thick all the way down to the event horizon for all reasonable mass accretion rates in which there is a thin accretion disk.

\begin{figure*}[t]
\begin{center}
\includegraphics[width=10cm,trim={0cm 3cm 7cm 1cm},clip]{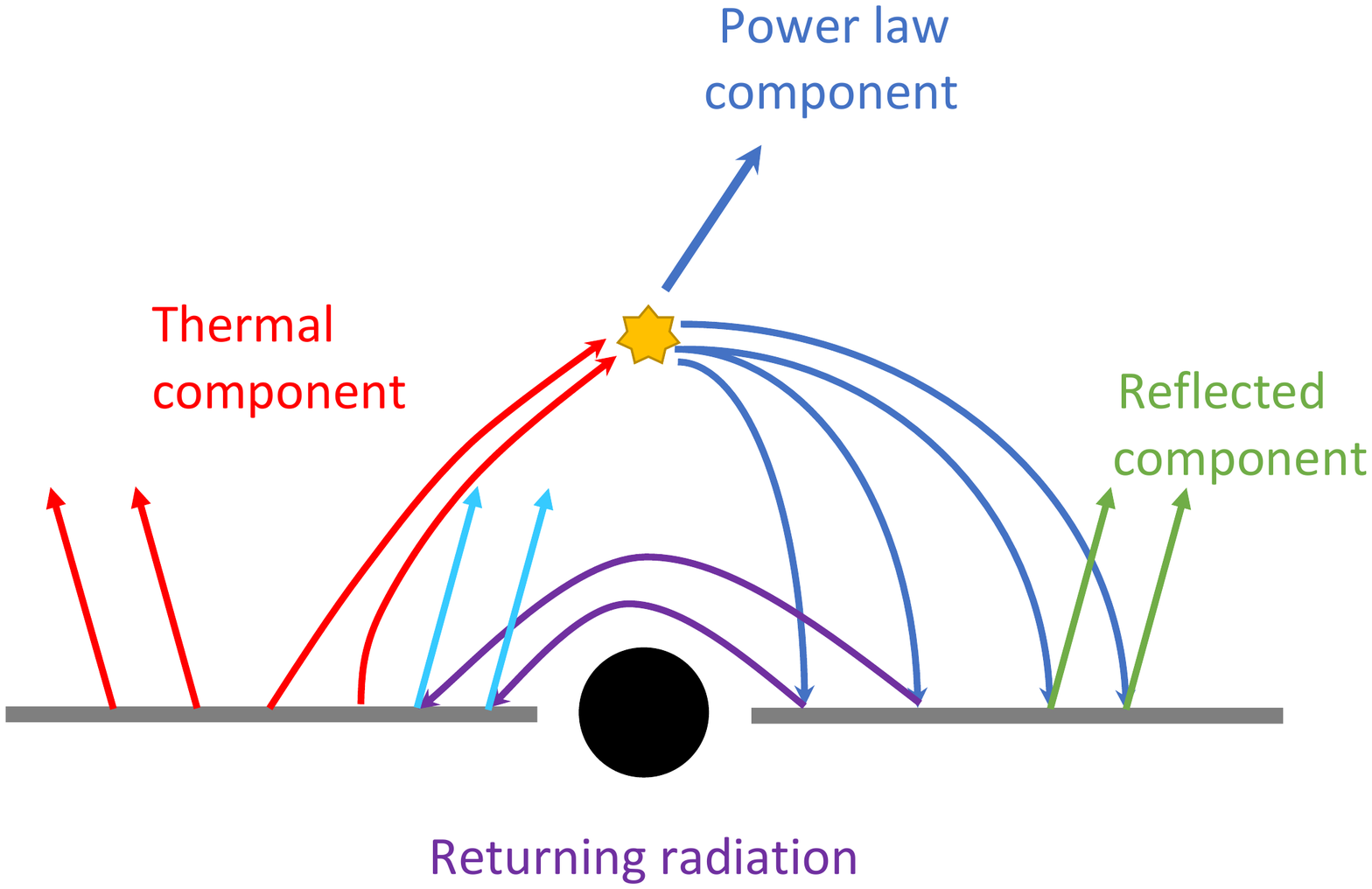}
\end{center}
\vspace{-0.8cm}
\caption{Components of the electromagnetic spectrum of an accreting black hole. The black hole (black circle) is surrounded by a geometrically thin and optically thick accretion disk (gray bars). The corona is represented by the yellow source above the black hole. Thermal photons from the disk (red arrows) inverse Compton scatter off free electrons in the corona, producing a power law component (blue arrows). The power law component illuminates the disk, producing a reflection component (green arrows). Some of the photons of the reflection component can return to the disk because of the strong light bending near the black hole: this is the returning radiation (violet arrows). The latter is reflected by the disk, producing a secondary reflection component (cyan arrows). Note that the reflection spectrum depends on the incident radiation, and the power law component and the returning radiation have different spectra.} \label{f-sketch}
\vspace{0.5cm}
\end{figure*}

The issue is instead more controversial in the case of the returning radiation, namely the radiation emitted by the disk and returning to the disk because of the strong light bending near the compact object (see Fig.~\ref{f-sketch}). For the reflection spectrum, we are not aware of any systematic study to figure out the modeling bias on the estimate of the parameter of an accreting black hole when the theoretical model does not include the calculation of the returning radiation. There are instead a few studies in the literature for the thermal component. The returning radiation can be ignored in the spectral analysis of the thermal component of the disk, as its effect can be reabsorbed into a higher mass accretion rate~\citep{2005ApJS..157..335L}. On the contrary, the returning radiation is crucial in the calculation of the polarization of the X-ray spectrum of black holes in the thermally-dominated state, since the scattered radiation is highly polarized~\citep{2009ApJ...701.1175S}. Note that in the case of the reflection spectrum, we cannot reabsorb the effect of the returning radiation in the emissivity profile of the disk, because the spectrum of the returning radiation of the reflection component has a reflection spectrum while the spectrum of the primary incident radiation from the corona is a power-law with an exponential high energy cut-off.

In the present paper, we want to evaluate the impact of the returning radiation on the reflection spectrum of an accretion disk. We simulate a set of observations, where in some simulations we include and in other simulations we do not include the returning radiation in the calculations. Then we fit the simulated observations with a relativistic reflection model that does not take the returning radiation into account; in this way we are able to estimate the modeling bias of the returning radiation when the theoretical model does not include the returning radiation. Our model is based on some important simplifications, in particular we assume that the material of the accretion disk is neutral. Our work is thus a preliminary study to figure out the actual relevance of the returning radiation in the measurement of the parameters of an accreting black hole, and additional work will be required to fully address this question.

The rest of the paper is organized as follows. In Section~\ref{r-rads}, we review previous studies on returning radiation in literature. In Section~\ref{s-mod}, we calculate some synthetic reflection spectra with and without returning radiation with the model employed in this study. In Section~\ref{s-sim}, we present our simulated observations and their analysis. We discuss our results in Section~\ref{s-con}.

%%%%%%%%%%%%%%%%%%%%%%%%%%%%%%%

\section{Returning radiations}\label{r-rads}

Fig.~\ref{f-sketch} shows an accreting black holes and the main components of its electromagnetic spectrum. The disk is geometrically thin and optically thick and any point of the disk is in thermal equilibrium. Every point on the surface of the disk has a blackbody-like spectrum. Since the temperature of the disk increases as gas falls into the gravitational well of the black hole, the total thermal spectrum of the disk is a multi-temperature blackbody-like spectrum. The emission from the inner edge of the accretion disk is normally peaked in the soft X-ray band for stellar-mass black holes and in the optical/UV bands for the supermassive ones. The corona is a hotter, say $\sim 100$~keV, often compact and optically thin, medium near the black hole. Thermal photons from the disk inverse Compton scatter off free electrons in the corona, acquiring a power law spectrum with an exponential high energy cut-off. The corona illuminates the disk, producing a reflection component. A fraction of the reflection radiation returns to the disk because of the strong light bending near the black hole. Such radiation illuminates the disk again, producing a secondary reflection component. Note that the spectrum of the reflection component depends on the spectrum of the incident radiation illuminating the disk. The radiation from the corona producing the {\it primary} reflection component has a power law spectrum. The returning radiation producing the {\it secondary} reflection component has a reflection spectrum produced by a power law spectrum.

\begin{figure*}[t]
\begin{center}
\includegraphics[type=pdf,ext=.pdf,read=.pdf,width=13.5cm]{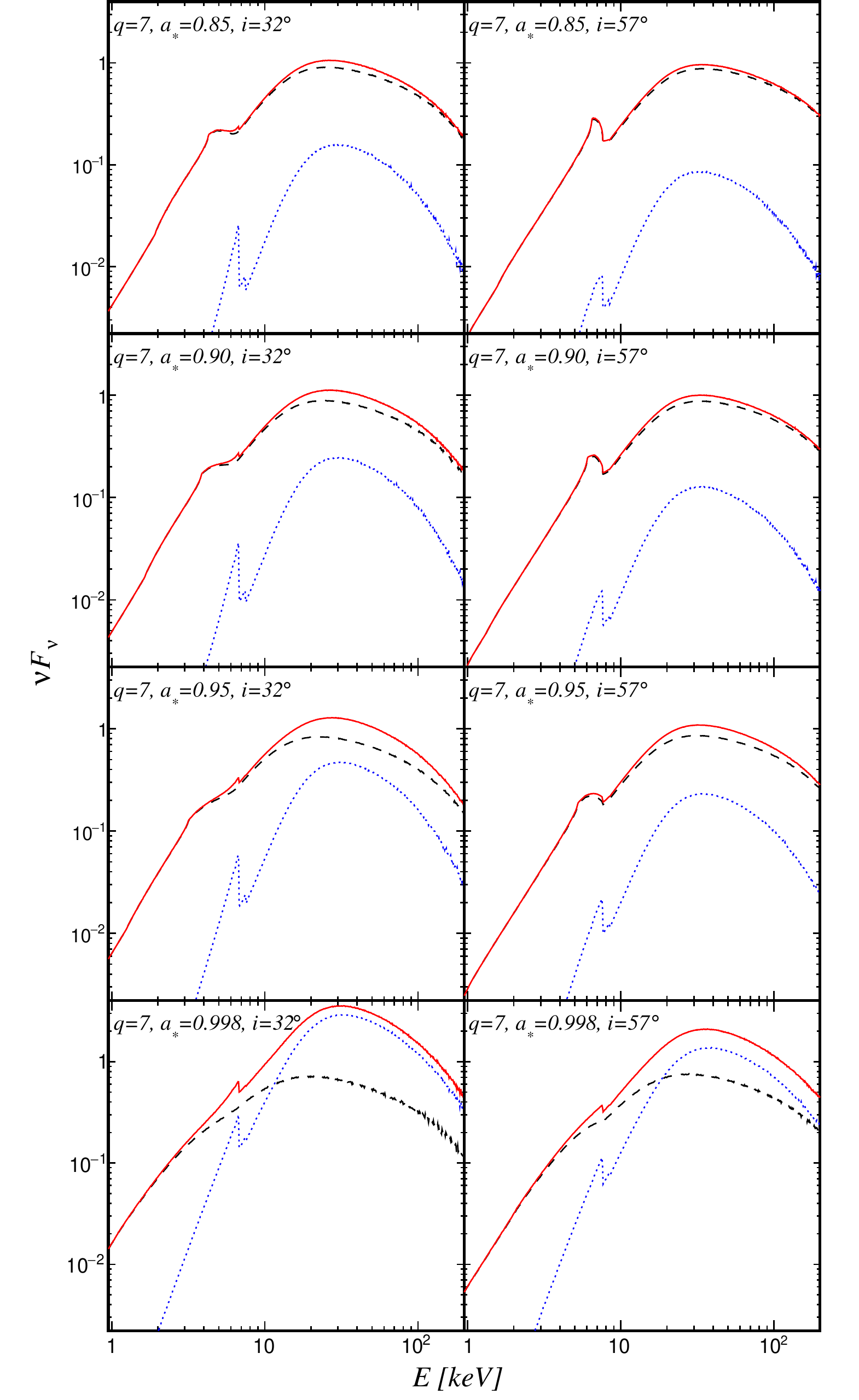}
\end{center}
%\vspace{-0.1cm}
\caption{Synthetic reflection spectra that include (red solid curves) and do not include (black dashed curves) the calculation of the returning radiation. Blue dotted curves represent the second-order reflection spectra. The black hole spin parameter is $a_* = 0.85$, 0.9, 0.95, and 0.998 (from top to bottom). The inclination angle of the disk is $i = 32^{\circ}$ (left panels) and $i = 57^{\circ}$ (right panels). The emissivity profile of the disk is described by a power law with emissivity index $q = 7$. In all cases the primary spectrum is a power-law with $\Gamma=2$, normalized at 1 keV.} \label{f-spec}
\end{figure*}

\begin{figure*}[t]
\begin{center}
\includegraphics[type=pdf,ext=.pdf,read=.pdf,width=13.5cm]{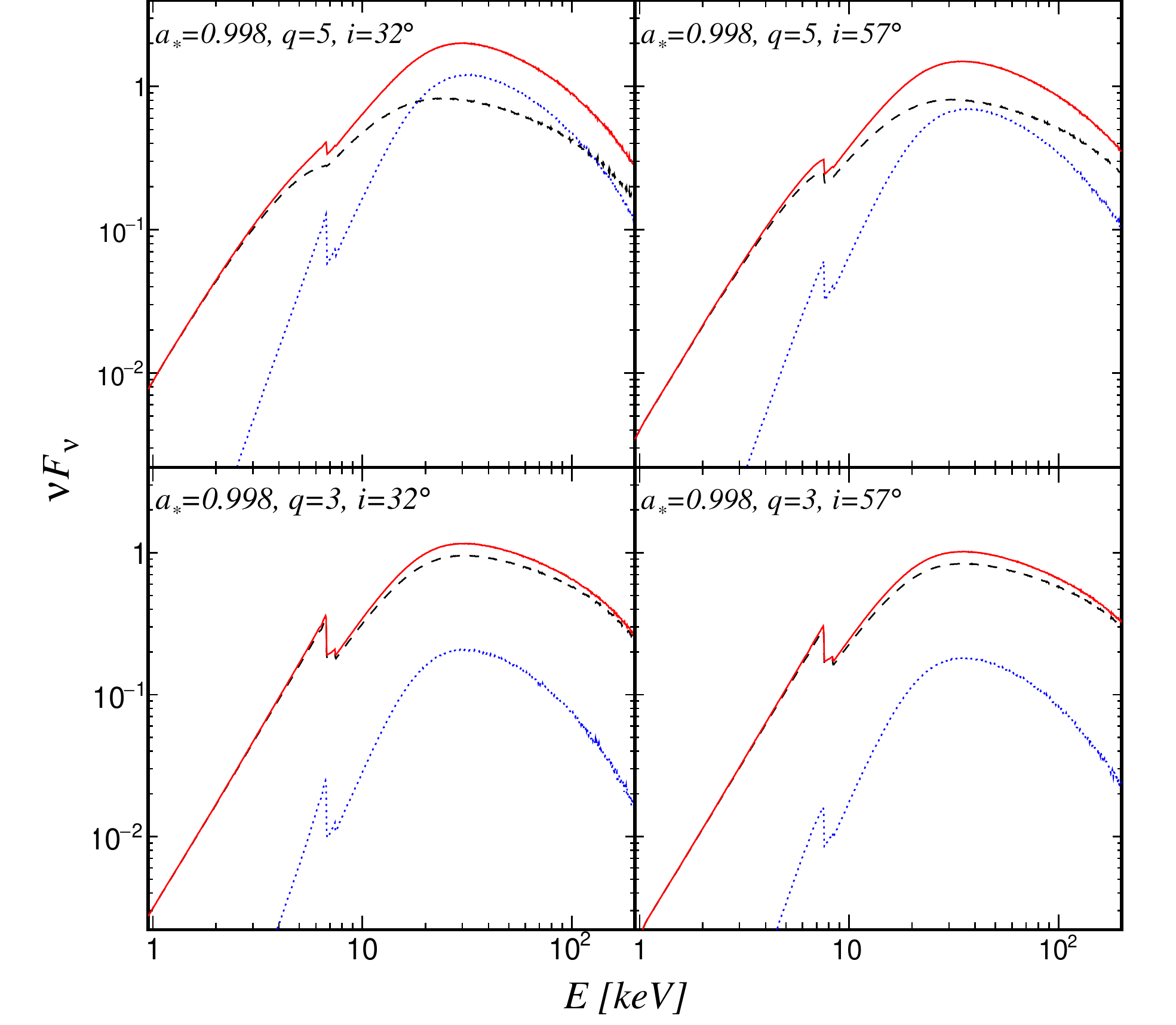}
\end{center}
%\vspace{-0.1cm}
\caption{As in Fig.~\ref{f-spec} but for an emissivity index $q = 5$ (top panels) and $q = 3$ (bottom panels). The black hole spin parameter is $a_* = 0.998$.  } \label{f-spec_q3_q5}
\end{figure*}

\citet{CT} was the first to include the calculation of the returning radiation into the spectrum of the thermal emission of an accretion disk around a black hole. He found that this effect can distort the observed spectrum but only for rapidly rotating black holes. More recent works have mostly considered the returning radiation in the context of the X-ray reflection. These studies were often limited to a specific range of parameters, and thus, different conclusions where found depending on whether the chosen parameters yielded a strong emission from the innermost disk. A significant fraction of radiation returns to the disk only if it comes from the inner few gravitational radii, because only in this region the light bending is sufficiently strong. For the emissivity profile of a standard accretion disk the contribution from that region is relatively weak and, therefore, the effect of the returning radiation on the reflected component is insignificant \citep{Dab1997}, but it can increase if the non-zero stress inner boundary condition is applied \citep{AK2000}. 

\begin{figure*}
\begin{center}
\includegraphics[type=pdf,ext=.pdf,read=.pdf,width=9cm]{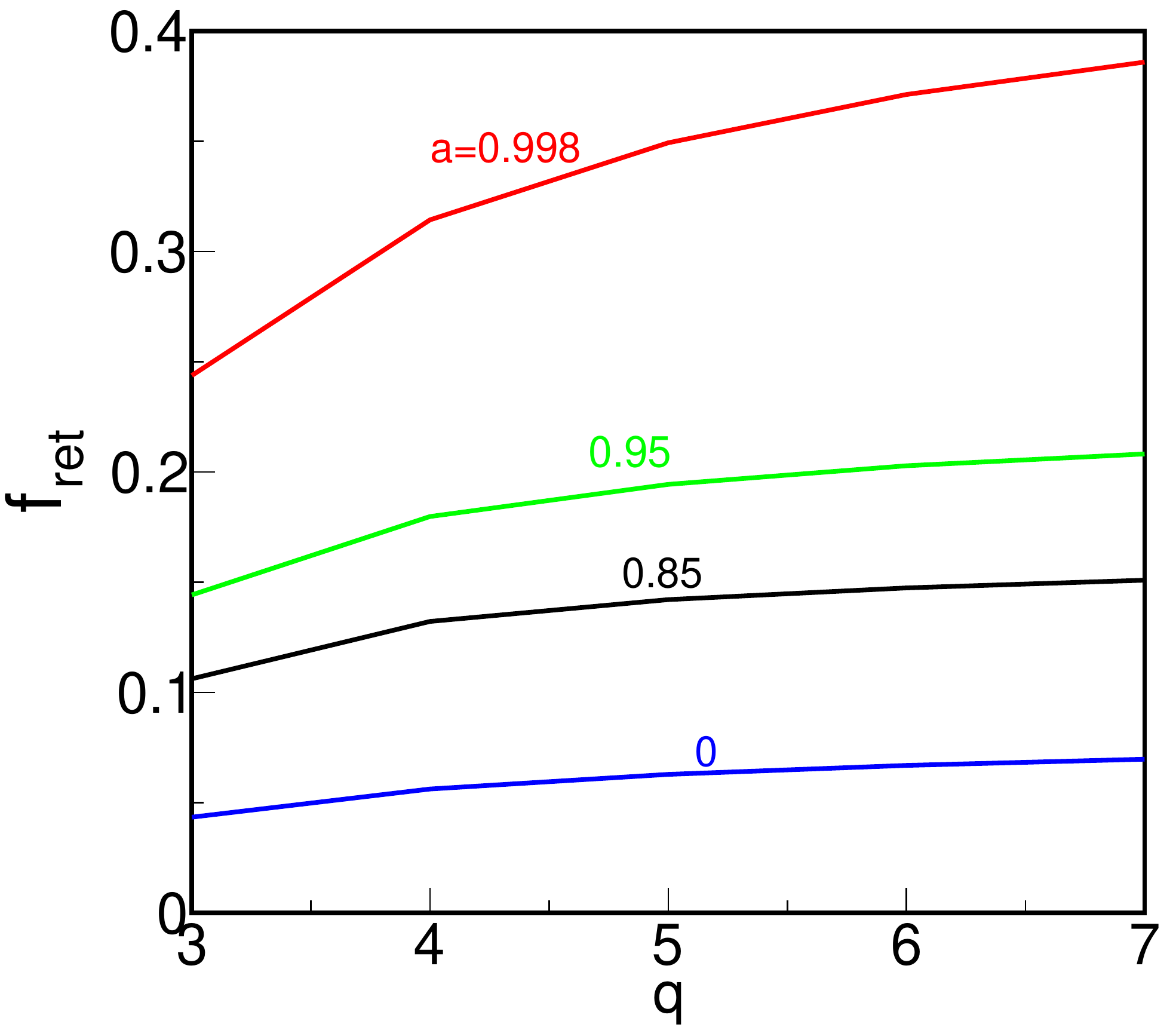}
\end{center}
%\vspace{-0.1cm}
\caption{Fraction of radiation returning to the disk for various values of the dimensionless spin parameter $a_{*}$ as a function of the emissivity index $q$ of the power-law irradiation profile. } \label{f-ret-frac}
\end{figure*}   

In the lamppost geometry, with the compact X-ray source on the symmetry axis of the disk, the importance of returning radiation depends on the height of the source, $h$. If the source is located relatively far from the black hole (e.g.\ at $h=10$~$R_{\rm g}$, where $R_{\rm g} = G_{\rm N}M/c^2$ is the gravitational radius, as assumed by \cite{WF2012}), reflection from the innermost disk is weak and the returning radiation has a negligible effect. If the X-ray source is located close to the black hole, at $h \la 2$~$R_{\rm g}$, the secondary reflection may outweigh the 1st-order one \citep{2016ApJ...821L...1N}.
A systematic exploration of this effect in the lamppost geometry is presented in \citet{2018MNRAS.477.4269N}, who considered the total re-emission of the incident radiation to make their results independent of the ionization state of the disk. They find that the 2nd-order re-emission is important only for high spin values, $a_* \ga 0.9$, low $h$, and when the disk is untruncated.
For angle-averaged emission, the 2nd-order re-emission is higher than the first order for $h \la 1.5$~$R_{\rm g}$. 
They also note that the dependence on the viewing angle, $i$, is crucial. In general, the returning radiation is more important for lower $i$. For face-on observers it can still give an important contribution up to a height of several $R_{\rm g}$.

\citet{sue2006} considered an X-ray source corotating with the underlying accretion disk. The azimuthal motion collimates radiation toward outer parts of the disk (which, thus, reduces irradiation of the innermost part), therefore, they found a rather modest returning radiation effect despite assuming a relatively small distance of the X-ray source from the black hole.

The comparison of reflection spectra including and neglecting the returning radiation in \citet{sue2006} and \citet{2016ApJ...821L...1N} shows that the strongest effect of the second order reflection occurs at $E \ga 10$ keV. This results, however, from their assumption that the reflection is neutral, in which case the reflected radiation is strongly attenuated by photo-absorption at $E \la 10$ keV.
The only relevant computation for an ionized reflection was presented in \citet{2002MNRAS.336..315R}. They did not consider the general-relativistic transfer. Instead they computed  multiple reflection from the same ionized medium, so their results can be used to infer the returning radiation effects in an ionized disk. They concluded that the net effect of multiple, ionized X-ray reflection is to steepen the soft X-ray part of the spectrum and strengthen the absorption and emission features. The overall effect of multiple reflection is to make the spectrum resemble a single reflection spectrum with a slightly higher ionization and significantly higher elemental abundances. The latter is particularly interesting, because it could explain the significantly super-solar abundance of iron found in many fitting results; we discuss this issue in Section \ref{s-con}. Here we note that the absorption and emission spectral features will be obviously strengthened by multiple reflection if photon energies are not changed between subsequent reflections, as assumed by \citet{2002MNRAS.336..315R}. However, this mechanism would be reduced if relativistic smearing between subsequent reflections is involved, which is typically the case for returning radiation. We also note that \citet{2002MNRAS.336..315R} based their conclusion on a qualitative spectral resemblence, without any quantitative assessment of the similarity of multiple-reflection and high Fe-abundance spectra.  

In this work we attempt to quantitatively evaluate the effect of returning radiation on parameters obtained from spectral fitting of the relativistic reflection models. We assume reflection to be neutral. Such neutral reflection has been sporadically reported \citep[e.g.][]{2012ApJ...755...88R}, but usually reflection in accreting systems is from ionized media. However, we can make an exact simulation only for the neutral case, which then dictated our choice of the model. Returning radiation distorts the relativistic reflection spectra through two effects: (i) radial redistribution of the irradiating flux, and (ii) contribution of reflection produced by radiation with energy distribution deviating from a power-law. Effect (i) is independent of the disk ionization state. Effect (ii) strongly depends on the ionization and, because the deviation of reflection from a power-law is stronger at lower ionization, we study here the maximum effect of the returning radiation on reflection, at least in the Fe K energy range. The soft X-ray range, where  the returning radiation effect may increase with ionization, is not considered in this work.

Bearing in mind the above noted results of previous works, here we focus on models with a disk extending down to the innermost stable circular orbit (as usually assumed in spectral fitting) and a steep radial emissivity, given by a power-law with the emissivity index $q=7$. Such a steep (or even steeper) emissivity from the inner disk was reported in many studies of AGNs and black hole binaries  \citep[e.g.][]{2012MNRAS.419..116F,2013MNRAS.429.2662B,2014ApJ...780...78T,2015ApJ...806..149K,2018ApJ...864...25G}. Steep profiles were also found to be common in a large sample of AGNs studied by \citet{2013MNRAS.428.2901W} when the reflection from the outer disk is faint enough that it does not significantly impact the observed spectrum, and on a statistical basis there is no need for a broken power-law. We also consider a few simulations with $q=5$ and $q=3$ to illustrate the impact of the emissivity index on the returning radiation, as well as a few cases more representative of the observed black holes with a broken power-law with $q_{\rm in} = 7$, $q_{\rm out} = 3$ and a breaking radius $r_{\rm br} = 3$ or 4~$R_{\rm g}$.

\section{The model}\label{s-mod}

In the present work, we employ the model developed by \citet{Nied2008}, below referred to as \texttt{reflection}, to calculate reflection spectra of thin accretion disks around Kerr black holes, either ignoring or including the returning radiation. This model combines a ray-tracing code, which calculates the direct radiation of the X-ray source and its reflection as seen by a distant observer, with a Monte Carlo code, which simulates the reflection process in the disk. Here we assume that the radial emissivity of reflection has a power-law form, and, thus, we do not compute the source-to-disk transfer and we simply generate photons in the disk frame with the probability distribution giving this radial profile. 
A large number of photons in the rest-frame of the disk are fired isotropically. Half are emitted away from the disk and contribute to the directly observed component, and the other half enter the disk. For those entering the disk we perform the Monte Carlo simulation of diffusion in cold matter, in which photons are subject to Compton scattering and photo-absorption.

Our treatment of Compton scattering follows the method of \citet{1983ASPRv...2..189P} and \citet{1984AcA....34..141G}, with scattering probability given by the differential Klein-Nishina cross section. We use photo-absorption opacities from \citet{1983ApJ...270..119M}. For photons with the initial energy $E_{\rm disk} > 7.1$~keV we also generate iron K fluorescent photons at 6.40 keV (K$\alpha$) and 7.06 keV (K$\beta$). We use the semi-analytic formula for the dependence of the probability of a fluorescent photon production and escape on the angle of incidence and the initial energy of \citet{1991MNRAS.249..352G}, with additional correction factors which allow us to reproduce the dependence of the iron K line intensity on the iron abundance, $A_{\rm Fe}$, and the emission angle, $\theta_{\rm em}$, of \texttt{xillver} \citep{2013ApJ...768..146G}. This, in particular, implies that $A_{\rm Fe}$ can be varied in the range $[0.5,10]$ (the same as in \texttt{xillver}).

For every Compton reflected and fluorescent photon, the equations of motion are solved numerically to determine whether the photon hits the surface of the disk in the equatorial plane, crosses the event horizon, or reaches the observer at infinity contributing to the observed 1st-order reflection. For photons hitting the disk, the 2nd-order reflection is simulated. The incidence angle and energy in the rest-frame of the disk are found and then the Monte Carlo code is used as described above. The transformation between the rest-frame parameters and the constants of motion is described in \citet{Nied2005}. The twice (or more times) reflected photons are traced again, and so, our simulated spectra take into account also higher order reflections; however, the contribution of more than twice reflected photons is negligible.

Synthetic reflection spectra without and with the returning radiation calculated with this model are shown in Fig.~\ref{f-spec} for the spin parameter $a_* = 0.85$, 0.9, 0.95, and 0.998 and for the viewing angle $i = 32^\circ$ and $57^\circ$. 
For lower values of the black hole spin parameter, the effect of the returning radiation is negligible. These spectra are calculated assuming that the emissivity profile of the disk is described by a power law with emissivity index $q=7$. The choice of a relatively high emissivity index is to enhance the emission from the inner part of the disk, where the probability that a photon returns to the disk is higher~\citep{AK2000}. To show the impact of the emissivity index on the contribution of the returning radiation in the total spectrum, we have simulated some spectra with a lower emissivity index. Fig.~\ref{f-spec_q3_q5} shows the spectra for an emissivity index $q = 5$ (top panels) and $q = 3$ (bottom panels), in the case of a black hole spin parameter $a_* = 0.998$. In all our simulations, the material of the disk is neutral and has solar iron abundance. The spectrum of the corona illuminating the disk is described by a power law with photon index $\Gamma = 2$. We clearly see that the contribution of the secondary reflection increases with increasing $a_*$ and decreasing $i$.

Fig.~\ref{f-ret-frac} shows the fraction of radiation returning to the disk as a function of the emissivity index $q$ for different values of the black hole spin parameter $a_*$. For high values of $a_*$, the inner edge of the disk is closer to the black hole, and thus increasing the value of $q$ has a stronger impact on the value of the fraction of radiation returning to the disk, with a saturation behavior for high values of the emissivity index $q$. For low values of the black hole spin parameter, the inner edge of the disk is relatively far that, even increasing the number of photons emitted from the inner edge, the fraction of radiation returning to the disk increases only moderately. For $a_* = 0$ ($r_{\rm ISCO} = 6$~$R_{\rm g}$) about 5\% of the radiation returns to the disk. As the black hole spin parameter increases and the inner edge of the disk gets closer to the black hole, the fraction of returning radiation increases up to almost 40\%.

Our results can be compared with those of \citet{AK2000}, who considered radiation emitted due to dissipation in the accretion disk and computed the returning-radiation fraction for two limiting cases, namely the classical Novikov-Thorne disk and the infinite efficiency disk (the latter with a strong torque exerted at ISCO). The fraction of returning radiation found in their Novikov-Thorne model is smaller than that in our $q=3$ models, by a factor of $\simeq 0.5$ and $\simeq 0.7$ for $a_*=0$ and $a_*=0.998$, respectively, which difference is fully understandable as the emissivity of the Novikov-Thorne disk is reduced compared to the $q=3$ profile in the innermost part, and the reduction is stronger for smaller $a_*$. Then, the emissivity in the infinite efficiency model of \citet{AK2000} has a very similar degree of central concentration to the $q=7$ profile\footnote{As can be estimated using Fig.~1 of \citet{AK2000}} and we note that the returning-radiation fractions in that model are also very similar, for any value of $a_*$, to ours for $q=7$. We then conclude that our results are consistent with \citet{AK2000}. On the other hand, for neither set of parameters we found the returning-radiation fraction as high as 75\%, which value was assumed by~\cite{2002MNRAS.336..315R}.

\citet{2020MNRAS.498.3302W} studied the effect of returning radiation on reflection spectra from an ionized disk's material in the case of a lamppost primary X-ray source. Their fractions of returning radiation for the lamppost height $h_{\rm lamppost} = 5~R_{\rm g}$ (their Tab.~1) match well our prediction of fraction of returning radiation for $q = 7$ shown in Fig.~\ref{f-ret-frac}. By lowering the lamppost height ($h_{\rm lamppost} = 1.5~R_{\rm g}$), the fraction of returning radiation can increase up to~47\% for a maximally rotating black hole (their Fig.~5). They estimated that the red wing of the iron K$\alpha$ line is enhanced relative to the continuum by~25$\%$, and the Compton hump about a factor of 3, in the presence of returning radiation.

%%%%%%%%%%%%%%%%%%%%%%%%%%%%%%%

\section{Simulations and fits}\label{s-sim}

In order to estimate the impact of the returning radiation on the measurement of the model parameters, we simulate some observations in which either we ignore the returning radiation or we include the returning radiation in the calculations. We then fit the simulated spectra with a reflection model that does not include the returning radiation, so that we can estimate the systematic uncertainties of the effect of the returning radiation. As X-ray mission, we choose \textsl{NICER}~\citep{nicer}, since it has a good energy resolution near the iron line and thus it is particularly suitable for precision measurements of relativistic features. Synthetic data are generated using the \texttt{fakeit} command in \texttt{xspec}~\citep{xspec} with the background, ancillary, and response matrix files of \textsl{NICER}. The \texttt{xspec} model to generate the simulations is \texttt{tbabs$\times$reflection}. \texttt{tbabs} describes the Galactic absorption~ \citep{wilms}. \texttt{reflection} calculates the reflection spectrum, see Section~\ref{s-mod}, for which we can choose either to ignore or to include the returning radiation. \texttt{reflection} also includes the observed primary spectrum, assumed here to be a power-law, and the relative normalization of the primary and reflected spectra corresponds to the isotropic emission of the primary radiation in the disk frame.

We run 16~observations for 4 possible values of the black hole spin parameter ($a_* = 0.85$, 0.9, 0.95, and 0.998), 2 possible values of the inclination angle of the disk ($i = 32^\circ$ or $57^\circ$), and either ignoring or including the returning radiation. In all these simulations, we assume that the intensity profile of the disk is described by a power law with emissivity index $q = 7$. We run 8~more simulations for the case $a_* = 0.998$ with emissivity index $q = 5$ and $3$. The choice of the value of the black hole spin for the simulations with lower emissivity index is determined by the fact that decreasing the emissivity index we reduce the fraction of radiation returning to the disk (see Fig.~\ref{f-ret-frac}) and therefore we compensate such an effect by choosing the highest value for the black hole spin. The table below summarizes our simulations, where we call NRR the simulations that ignore the returning radiation and WRR those that include the returning radiation in their calculations of the reflection spectrum: 

\vspace{0.3cm}

\begin{center}
\begin{tabular}{c|c|c|c|c}
\hline\hline
\hspace{0.3cm} $a_*$ \hspace{0.3cm} & \hspace{0.3cm} $i$ \hspace{0.3cm} & \hspace{0.3cm} $q$ \hspace{0.3cm} & \multicolumn{2}{c}{Returning Radiation} \\
& & \hspace{0.7cm} & \hspace{0.7cm}  No \hspace{0.7cm} & \hspace{0.7cm} Yes \hspace{0.7cm} \\
\hline
0.85 & $32^\circ$ & {7} &NRR1 & WRR1 \\
0.9 & $32^\circ$ &{7} &NRR2 & WRR2 \\
0.95 & $32^\circ$ &{7} & NRR3 & WRR3 \\
0.998 & $32^\circ$ & {7} & NRR4 & WRR4 \\
 {0.998} &  {$32^\circ$} &  {5} & {NRR4.1} &{WRR4.1} \\
{0.998} & {$32^\circ$} & {3} &{NRR4.2} &  {WRR4.2} \\
\hline
0.85 & $57^\circ$ &{7} & NRR5 & WRR5 \\
0.9 & $57^\circ$ & {7} & NRR6 & WRR6 \\
0.95 & $57^\circ$ & {7} & NRR7 & WRR7 \\
0.998 & $57^\circ$ & {7} &NRR8 & WRR8 \\
{0.998} & {$57^\circ$} & {5} & {NRR8.1} & {WRR8.1} \\
 {0.998} &  {$57^\circ$} & {3} & {NRR8.2} & {WRR8.2} \\
\hline\hline
\end{tabular}
\end{center}

\vspace{0.3cm}

We run 12 more observations for the black hole spin parameter $a_* = 0.9$ and $0.998$ and observer's inclination angle $i = 32^{\circ}$ and $57^{\circ}$, either excluding or including the returning radiation. In contrast to previous simulations, now the disk's irradiation profile is described by a broken power-law with the inner emissivity index $q_{\rm in} = 7$,  the outer emissivity index $q_{\rm out} = 3$, and the breaking radius $r_{\rm br} = 3~R_{\rm g}$ or $4~R_{\rm g}$. These emissivity indices and the breaking radii are somewhat realistic and close to what has been measured for several sources \citep[e.g.][]{2012MNRAS.419..116F,2013MNRAS.429.2662B,2014ApJ...780...78T,2015ApJ...806..149K,2018ApJ...864...25G}. We do not consider the input spin parameter less than 0.9 in our broken power-law setup because, at a lower spin parameter, the impact of returning radiation is not significant on the reflection spectra (see Fig.~\ref{f-ret-frac}). The synthetic data of these simulations are analyzed with the broken power-law mode. The table below summarizes our simulations -- including and excluding the returning radiation -- for a broken power-law emissivity profile of the accretion disk.  

\begin{center}
\begin{tabular}{c|c|c|c|c|c|c}
\hline\hline
\hspace{0.3cm} $a_*$ \hspace{0.3cm} & \hspace{0.3cm} $i$ \hspace{0.3cm} & \hspace{0.3cm} $q_{\rm in}$ \hspace{0.3cm} &  \hspace{0.3cm} $q_{\rm out}$ \hspace{0.3cm} &  \hspace{0.3cm} $r_{\rm br}$ \hspace{0.3cm} & \multicolumn{2}{c}{Returning Radiation} \\
& & \hspace{0.7cm} & \hspace{0.7cm} & \hspace{0.7cm} & \hspace{0.7cm} No \hspace{0.7cm} & \hspace{0.7cm} Yes \hspace{0.7cm} \\
\hline
0.9 & $32^\circ$ &{7}& {3} & {4} & NRR9 & WRR9 \\
0.998 & $32^\circ$ &{7}& {3} & {4} & NRR10 & WRR10 \\
0.998 & $32^\circ$ &{7}& {3} & {3} & NRR11 & WRR11 \\
\hline
0.9 & $57^\circ$ &{7}& {3} & {4} & NRR12 & WRR12 \\
0.998 & $57^\circ$ &{7}& {3} & {4} & NRR13 & WRR13 \\
0.998 & $57^\circ$ &{7}& {3} & {3} & NRR14 & WRR14 \\
\hline\hline
\end{tabular}
\end{center}

\vspace{0.3cm}

The iron abundance of the disk is set to the solar value ($A_{\rm Fe} = 1$), and the power law spectrum from the corona has photon index $\Gamma = 2$. These are the input parameters of the synthetic spectra calculated in the previous section. We simulate the observation of a bright Galactic X-ray binary and we set the photon flux to $4 \cdot 10^{-9}$~erg~cm$^{-2}$~s$^{-1}$ in the energy range 0.2-12~keV. We assume an exposure time of 300~ks, which leads to having about 100~million photons in the 0.2-12~keV band. While \textsl{NICER} cannot observe a source for such a long time, in principle we can imagine combining several consecutive observations of the same source and, regardless, our aim is to consider higher quality data than those available today.

\texttt{reflection} is too slow to fit the simulated data. The analysis of the simulated data is thus done with an \texttt{xspec} model that mostly follows the \texttt{reflkerr} model of \citet{reflkerr}, except for using \texttt{pexrav} as rest-frame reflection model. The general relativistic transfer of radiation is the same ray-tracing code as used in \texttt{reflection}, therefore, the relativistic smearing in both these models is exactly the same. The \texttt{pexrav} model \citep{1995MNRAS.273..837M} involves  an exact description of Compton reflection through Green's functions, which were computed with a Monte Carlo code similar to that used here in \texttt{reflection}. It also uses the same abundances and atomic cross sections as those in the \texttt{reflection} Monte Carlo simulation. Then, the \texttt{pexrav} reflection spectra precisely match  the spectra of the 1st-order reflection in the disk frame simulated with \texttt{reflection}. \texttt{xillver}, which is originally used in \texttt{reflkerr} for rest-frame reflection in the soft X-ray range, adopts the solar abundances of \citet{1998SSRv...85..161G}, while our Monte Carlo code uses those of \citet{1983ApJ...270..119M}.
For a weakly ionized reflector, the difference of the abundance models is relatively unimportant, but still the small differences of the reflection spectra \citep[see e.g.\ Figures 15 and 20 in][]{2013ApJ...768..146G} could possibly bias our result. \texttt{pexrav} computes the photo-absorbed reflection without any fluorescent lines. In our implementation we add to it the intrinsically narrow lines at 6.40 keV and 7.06 keV, representing the neutral Fe K$\alpha$ and $\beta$ lines, using the same scaling of their normalization with  $A_{\rm Fe}$, $\theta_{\rm em}$ and $\Gamma$ which underlies the fluorescence probability in our Monte Carlo simulations. The relative normalization of the primary and reflected component is scaled in \texttt{reflkerr} with the reflection fraction parameter, $\mathcal{R}$, where $\mathcal{R}=1$ corresponds to the isotropic primary emission in the disk frame, i.e.\ the same as assumed in our \texttt{reflection} simulation.

The simulated data are thus fitted with the \texttt{xspec} model \texttt{tbabs$\times$reflkerr}. In \texttt{tbabs} we freeze the hydrogen column density to its input value. \texttt{reflkerr} describes both the coronal spectrum and the reflection spectrum of a neutral disk in Kerr spacetime. The primary spectrum in this model is an e-folded power-law. The \texttt{reflection} spectra are simulated assuming a simple power-law, so to fit them we fix the folding energy at $E_{\rm cut} = 100$~MeV. Any effects related with more realistic values of $E_{\rm cut}$ cannot be measured with the low energy data of \textsl{NICER}. All other model parameters of \texttt{reflkerr} are free in the fit. We fit the simulated data in the energy range 2-10~keV, since for a neutral reflection the impact of the returning radiation is negligible below 2~keV and, on the contrary, the higher statistics at low energy tends to drive the fit and does not permit us to estimate the modeling bias induced by the returning radiation. We assume radial emissivity given by a single power-law, but in the case when it does not give a good fit (WRR4) we try also a broken power-law emissivity.

\begin{figure*}[t]
\vspace{0.5cm}
\begin{center}
\includegraphics[type=pdf,ext=.pdf,read=.pdf,width=9cm]{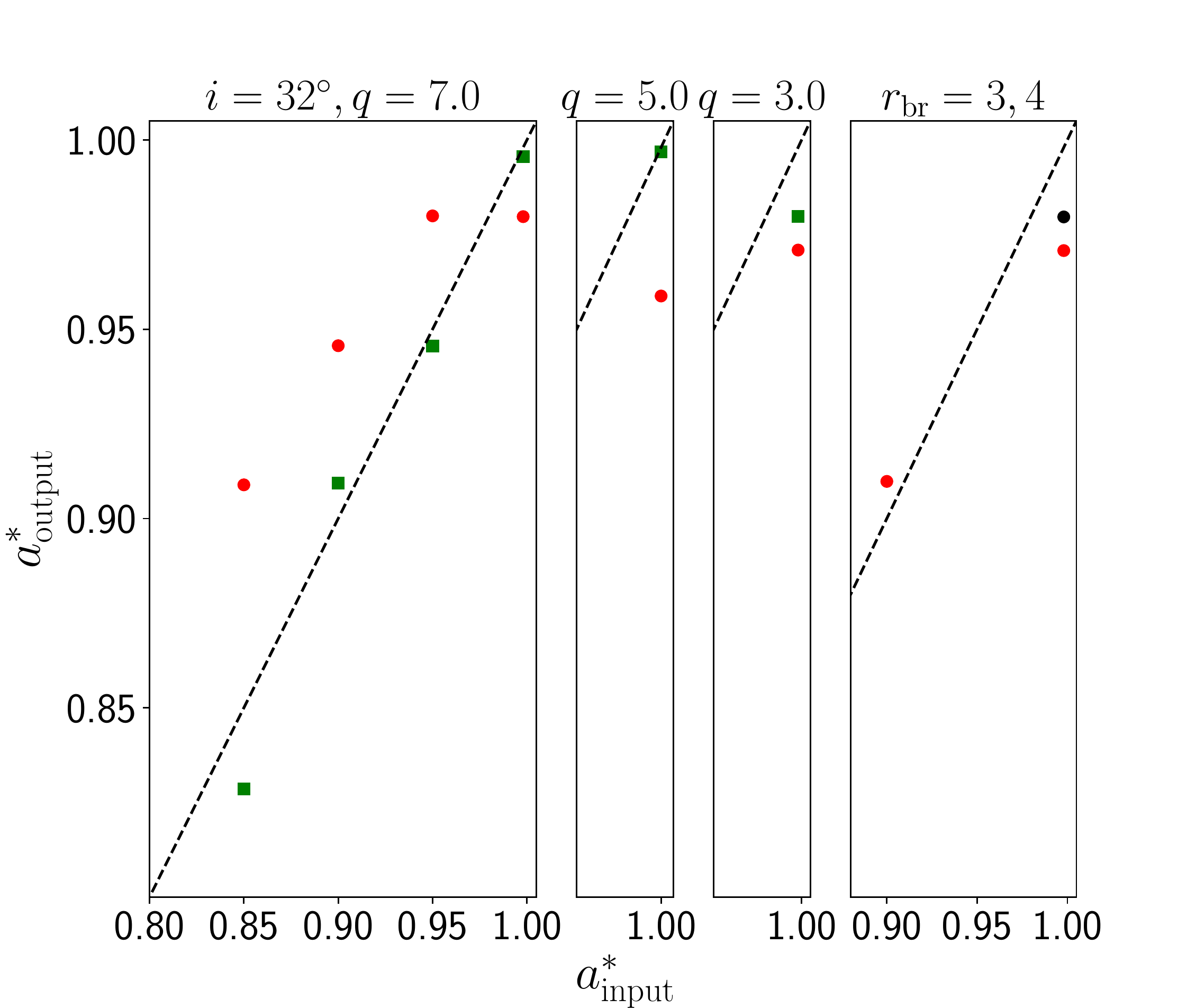}
\hspace{-0.2cm}
\includegraphics[type=pdf,ext=.pdf,read=.pdf,width=9cm]{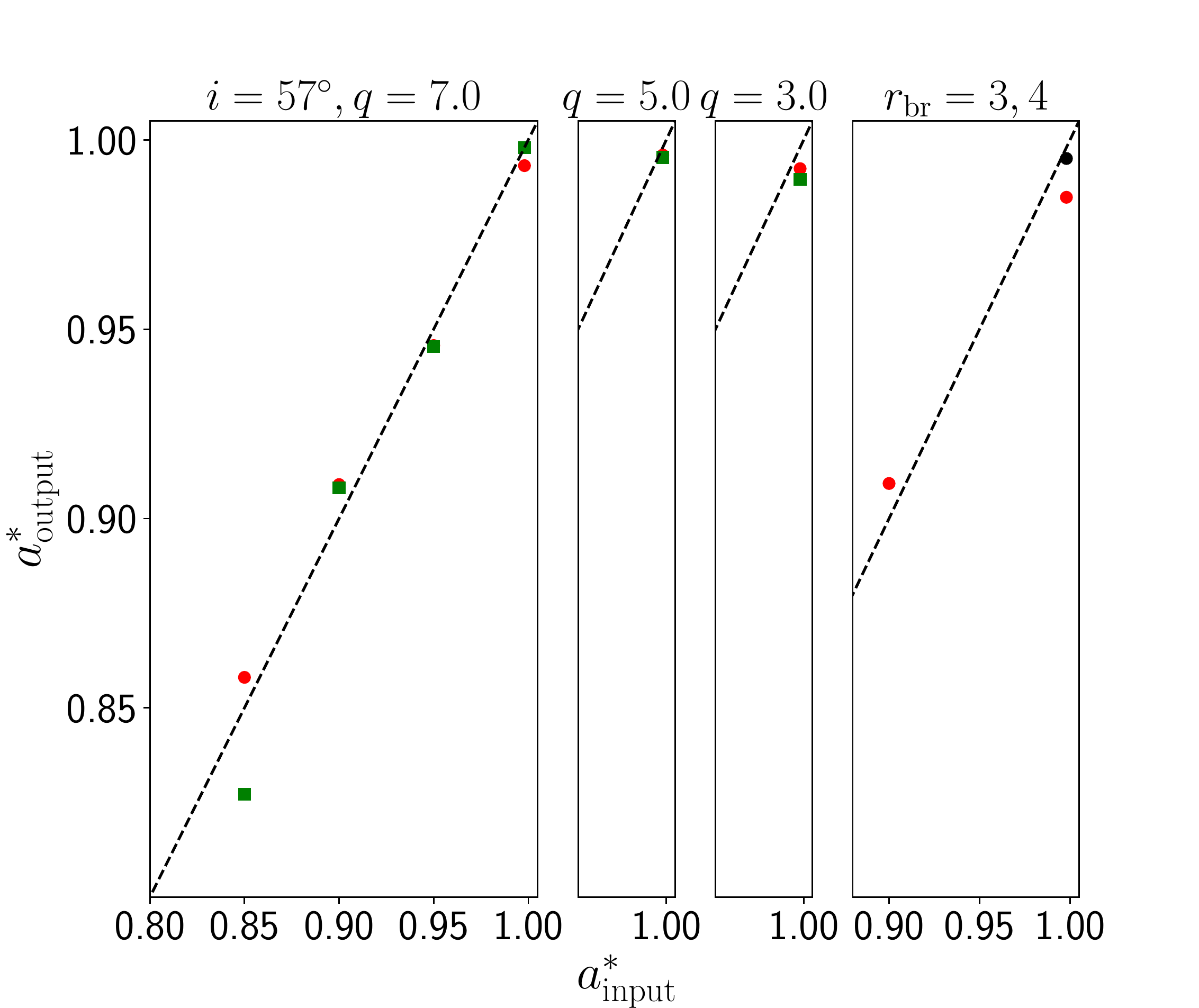}
\end{center}
\vspace{-0.1cm}
\caption{Input spin parameter of the simulations vs best-fit spin parameter inferred from the spectral analysis. The red circles  show the results from the simulations that include the returning radiation. The green squares show the results from the simulations that do not include the returning radiation. The black and red circles in the rightmost panel show the results from the simulations (with the returning radiation) with a broken power-law input emissivity profile with breaking radii, respectively, $r_{\rm br} = 3~R_{\rm g}$ and $4~R_{\rm g}$. \label{f-spins}}
%\end{figure*}
%%%%%%%%%%%%%%%%%%%%%%%%%%%%%%%%%%%%%%%%%%
%%%%%%%%%%%%%%%%%%%% incl. angle %%%%%%%%%%%%%%%%%%%
%\begin{figure*}[t]
\vspace{0.4cm}
\begin{center}
\includegraphics[type=pdf,ext=.pdf,read=.pdf,width=9cm]{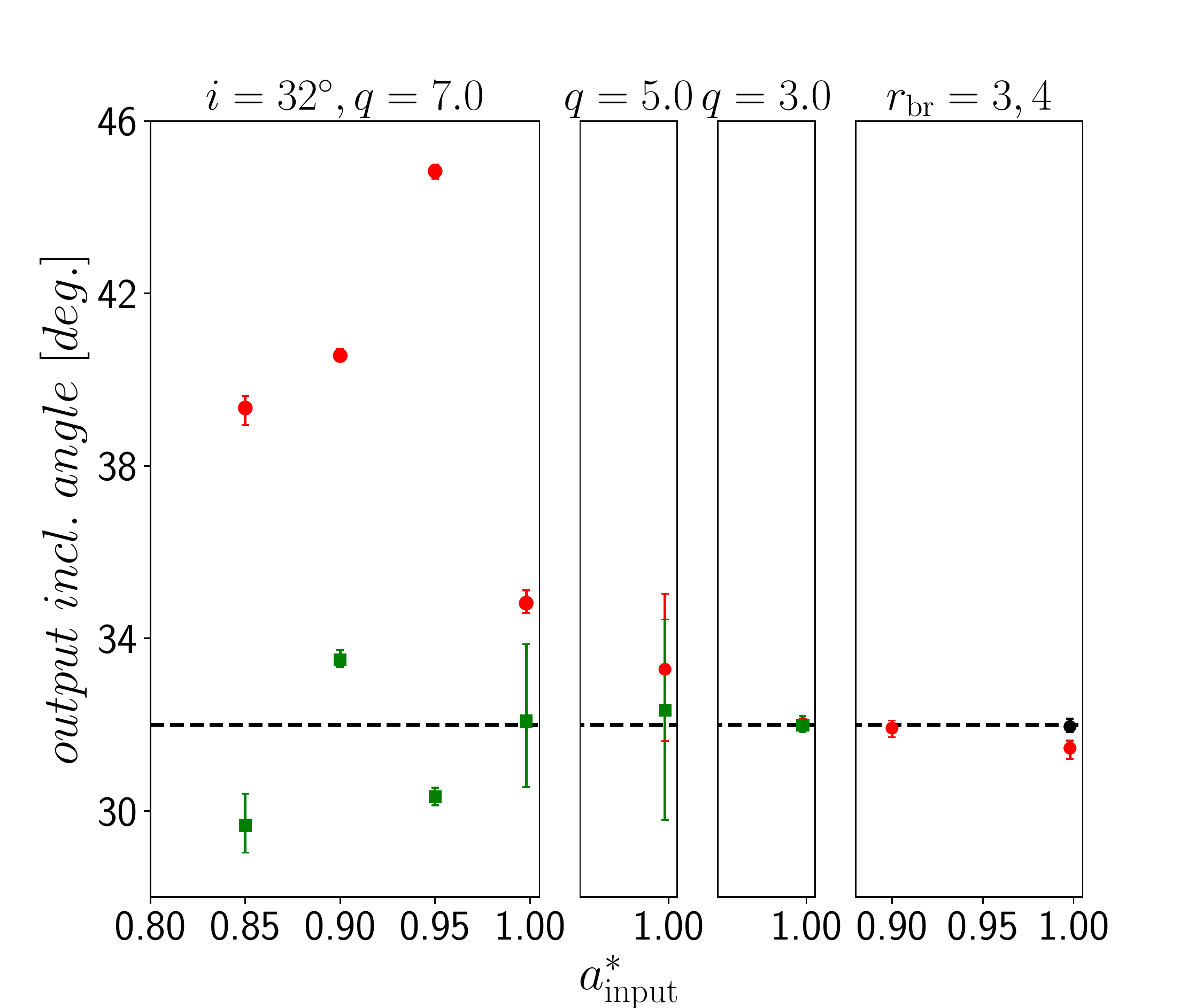}
\hspace{-0.5cm}
\includegraphics[type=pdf,ext=.pdf,read=.pdf,width=9cm]{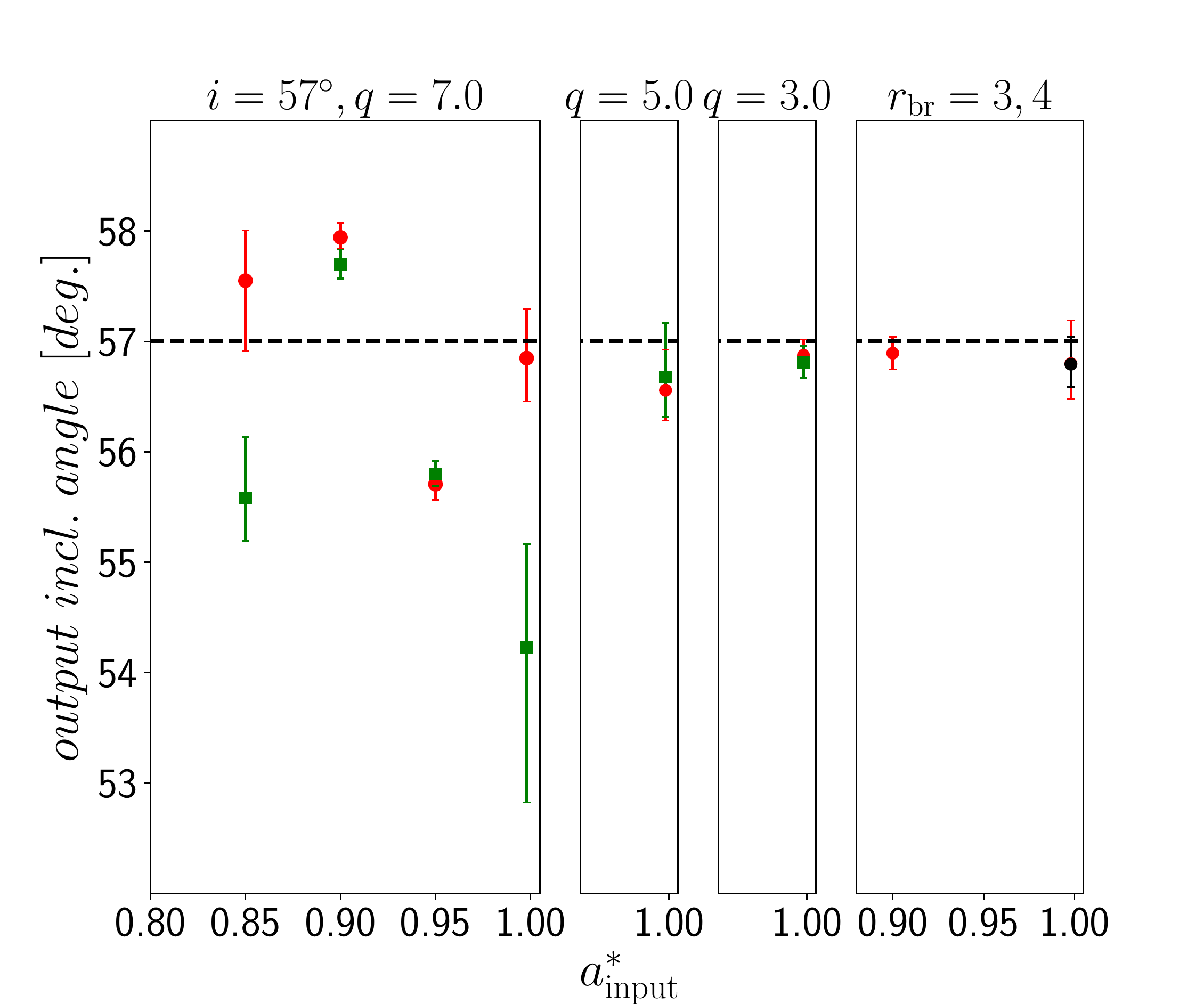}
\end{center}
\vspace{-0.1cm}
\caption{Input spin parameter of the simulations vs best-fit inclination angle inferred from the spectral analysis. The red circles show the results from the simulations that include the returning radiation. The green squares show the results from the simulations that do not include the returning radiation. The black and red circles in the rightmost panel show the results from the simulations (with the returning radiation) with a broken power-law input emissivity profile with breaking radii, respectively, $r_{\rm br} = 3~R_{\rm g}$ and $4~R_{\rm g}$.  \label{f-angle}} 
\vspace{0.5cm}
\end{figure*}

\begin{figure*}[t]
\vspace{0.5cm}
\begin{center}
\includegraphics[type=pdf,ext=.pdf,read=.pdf,width=9cm]{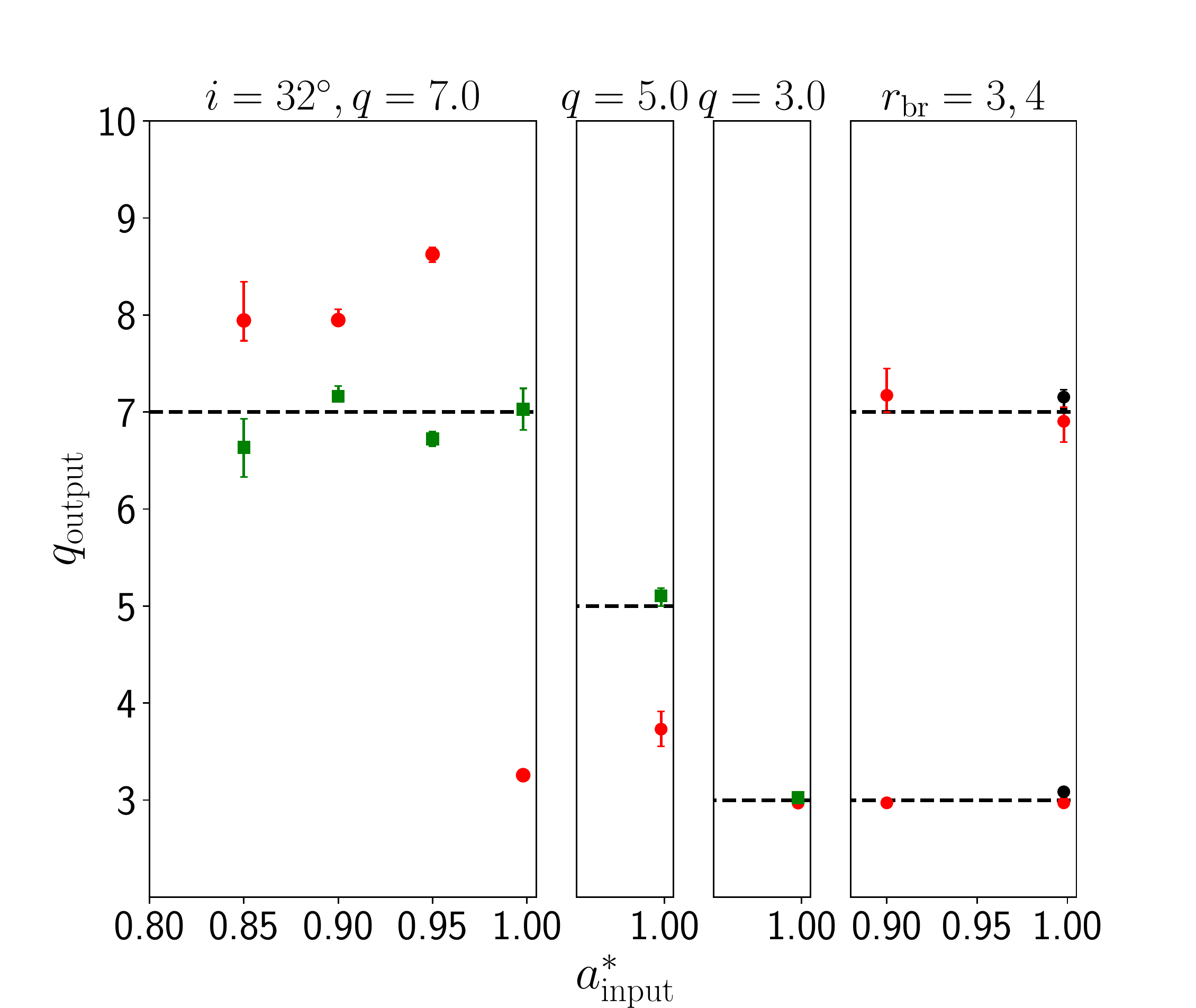}
\hspace{-0.5cm}
\includegraphics[type=pdf,ext=.pdf,read=.pdf,width=9cm]{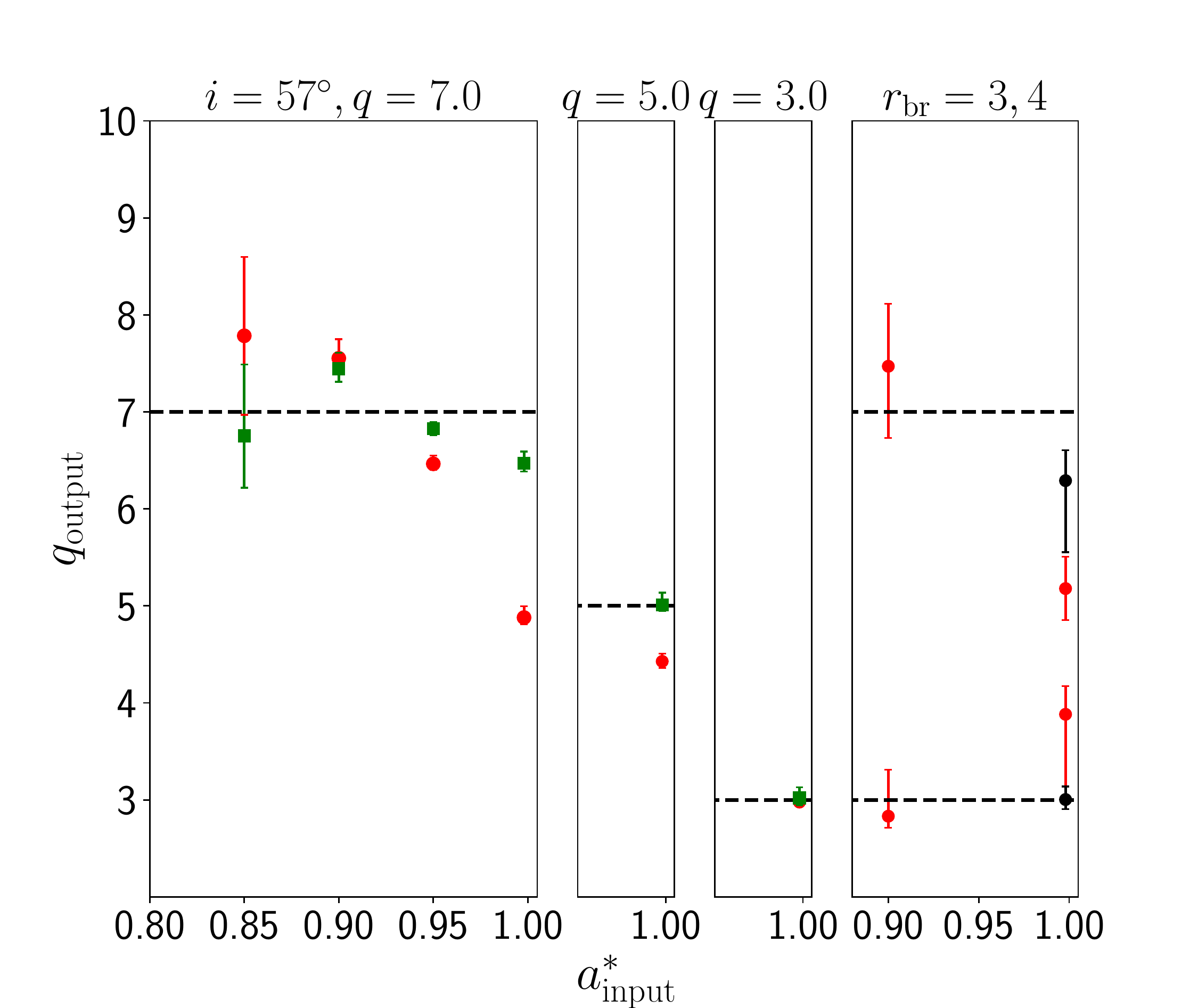}
\end{center}
\vspace{-0.1cm}
\caption{Input spin parameter of the simulations vs best-fit emissivity index inferred from the spectral analysis. The red circles  show the results from the simulations that include the returning radiation. The green squares show the results from the simulations that do not include the returning radiation.The black and red circles in the rightmost panel show the results from the simulations (with the returning radiation) with a broken power-law input emissivity profile with breaking radii, respectively, $r_{\rm br} = 3~R_{\rm g}$ and $4~R_{\rm g}$. The upper and lower circles show the results of inner and outer emissivity indices, respectively. \label{f-em-index}}
%\end{figure*}
%%%%%%%%%%%%%%%%%%%%%%%%%%%%%%%%%%%%%%%%%%
%%%%%%%%%%%%%%%%%%%%     Afe    %%%%%%%%%%%%%%%%%%%
%\begin{figure*}[t]
\vspace{0.4cm}
\begin{center}
\includegraphics[type=pdf,ext=.pdf,read=.pdf,width=9cm]{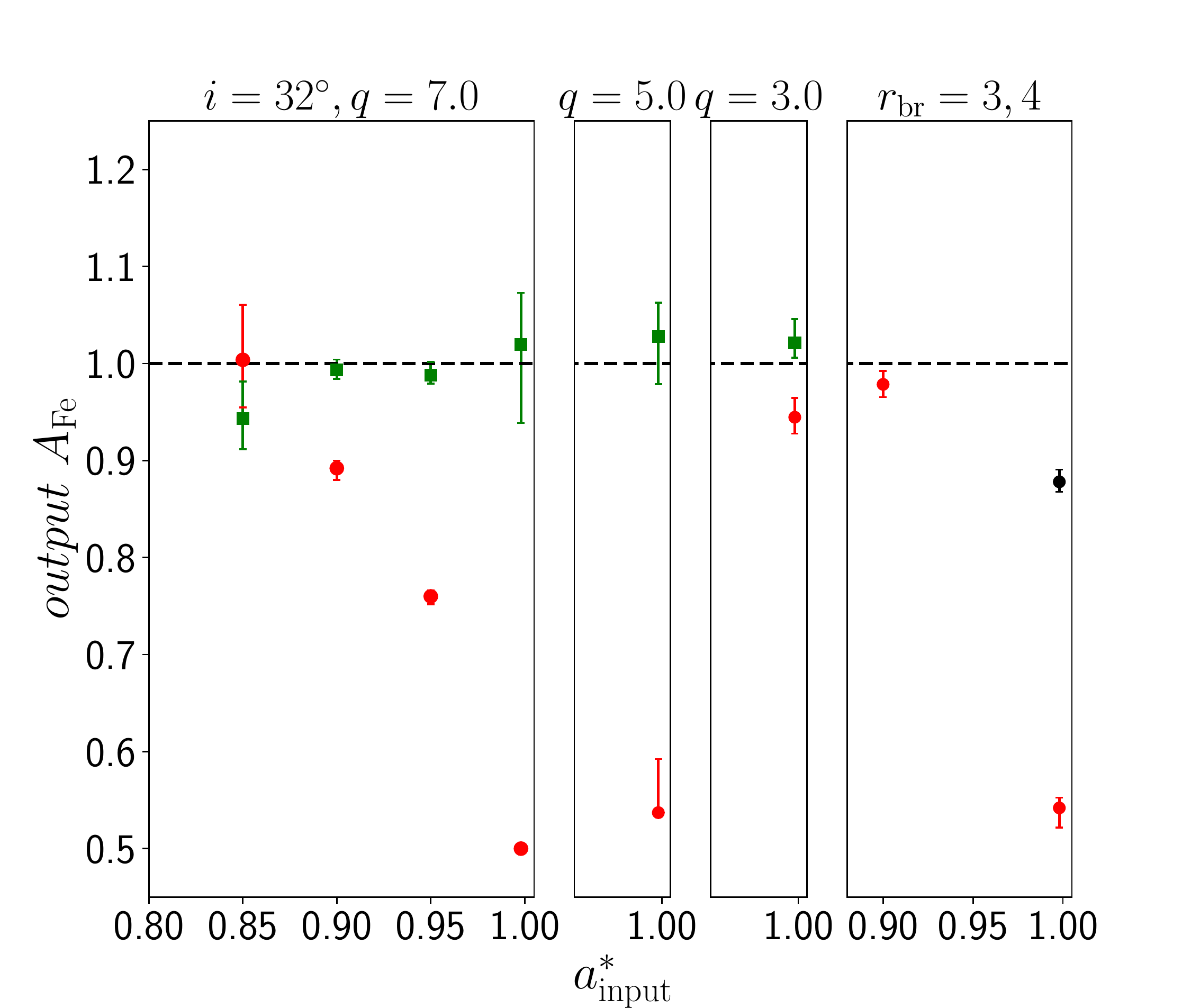}
\hspace{-0.5cm}
\includegraphics[type=pdf,ext=.pdf,read=.pdf,width=9cm]{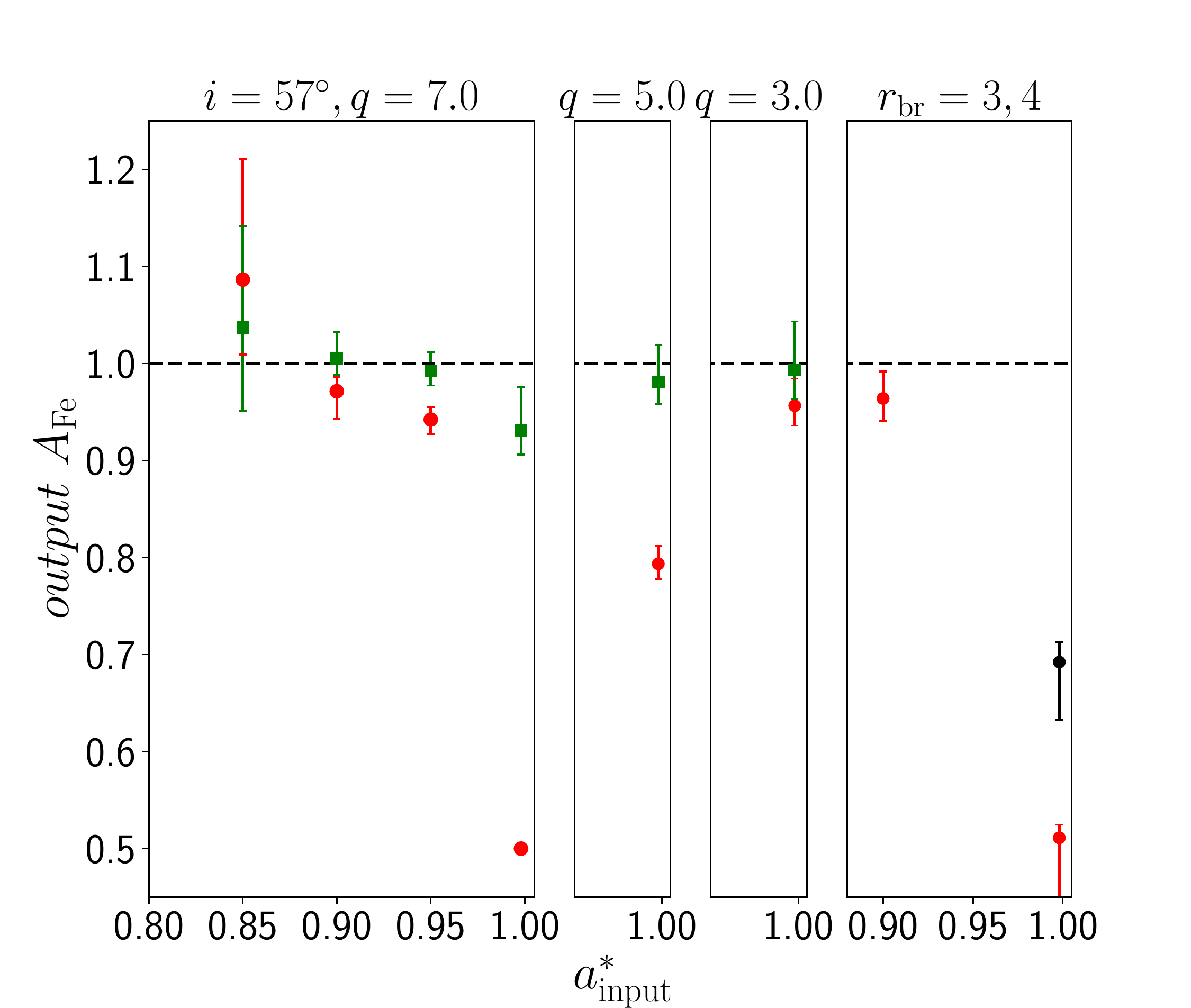}
\end{center}
\vspace{-0.1cm}
\caption{Input spin parameter of the simulations vs best-fit iron abundance inferred from the spectral analysis. The red circles show the results from the simulations that include the returning radiation. The green squares  show the results from the simulations that do not include the returning radiation. The black and red circles in the rightmost panel show the results from the simulations (with the returning radiation) with a broken-power law input emissivity profile with breaking radii, respectively, $r_{\rm br} =3~R_{\rm g}$ and $4~R_{\rm g}$.  \label{f-afe}}
\end{figure*}

\begin{figure*}[t]
\begin{center}
\vspace{0.5cm}
\includegraphics[type=pdf,ext=.pdf,read=.pdf,width=9cm,height=9cm]{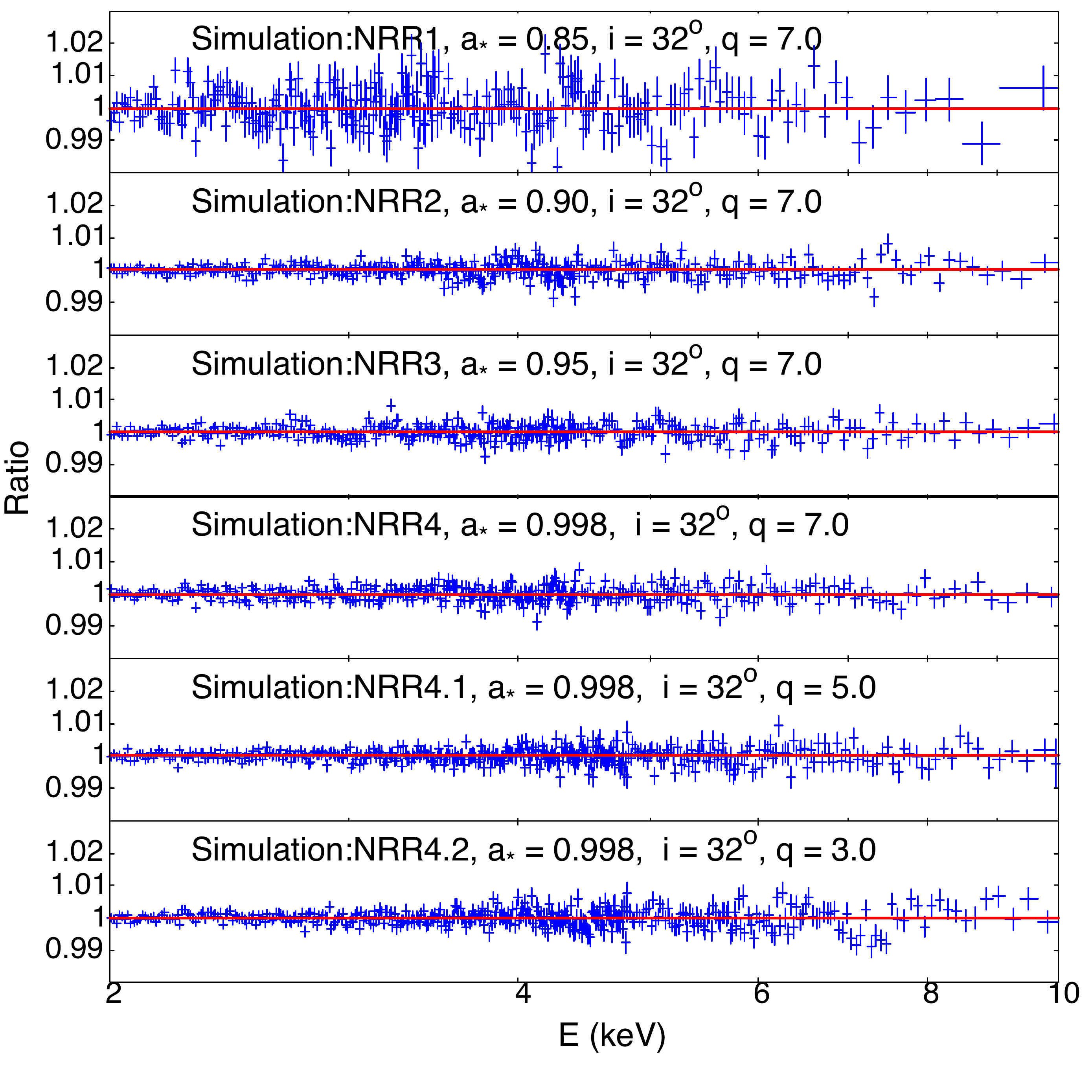}
\hspace{-0.2cm}
\vspace{-0.2cm}
\includegraphics[type=pdf,ext=.pdf,read=.pdf,width=9cm,height=9cm]{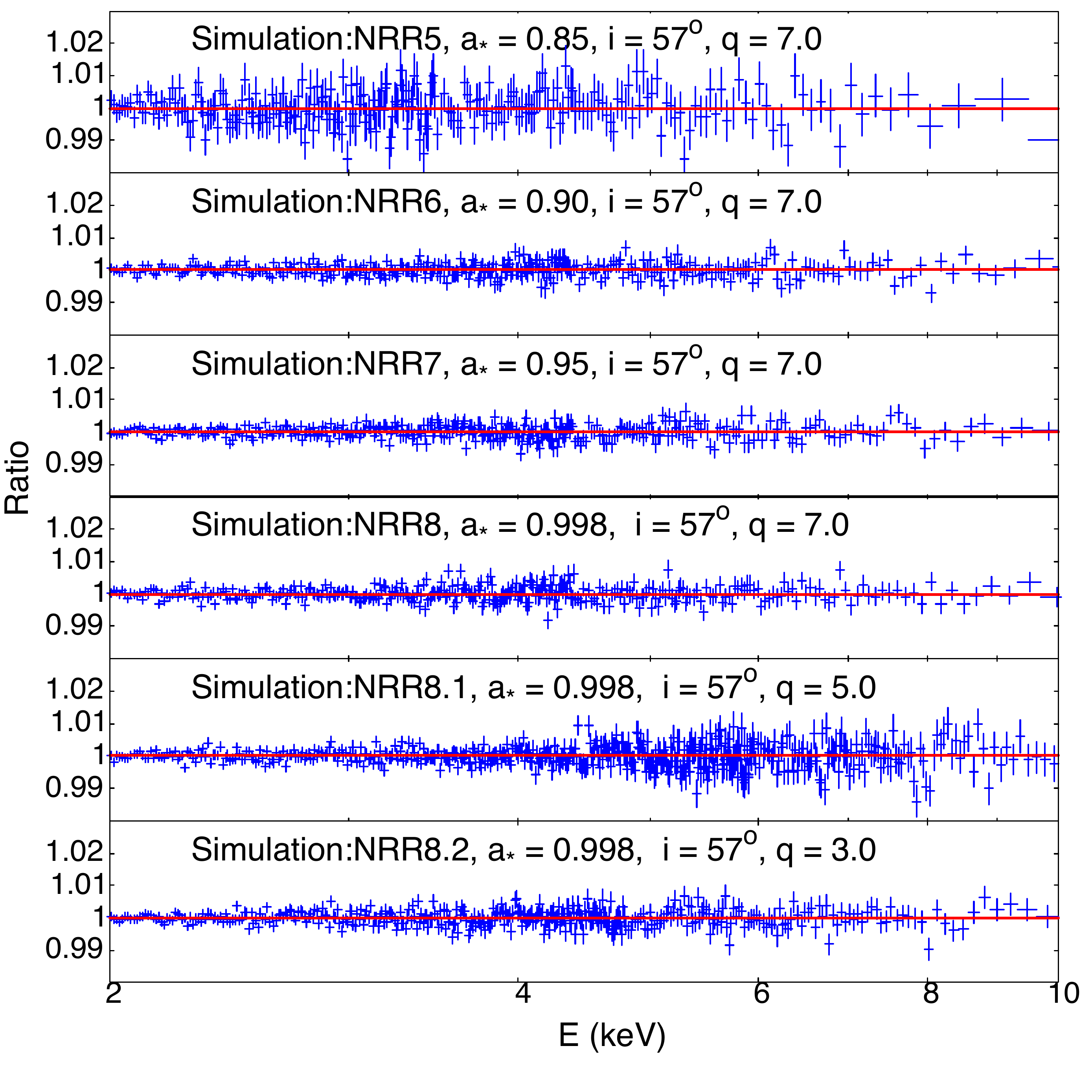}
%\vspace{0.5cm}
\includegraphics[type=pdf,ext=.pdf,read=.pdf,width=9cm,height=9cm]{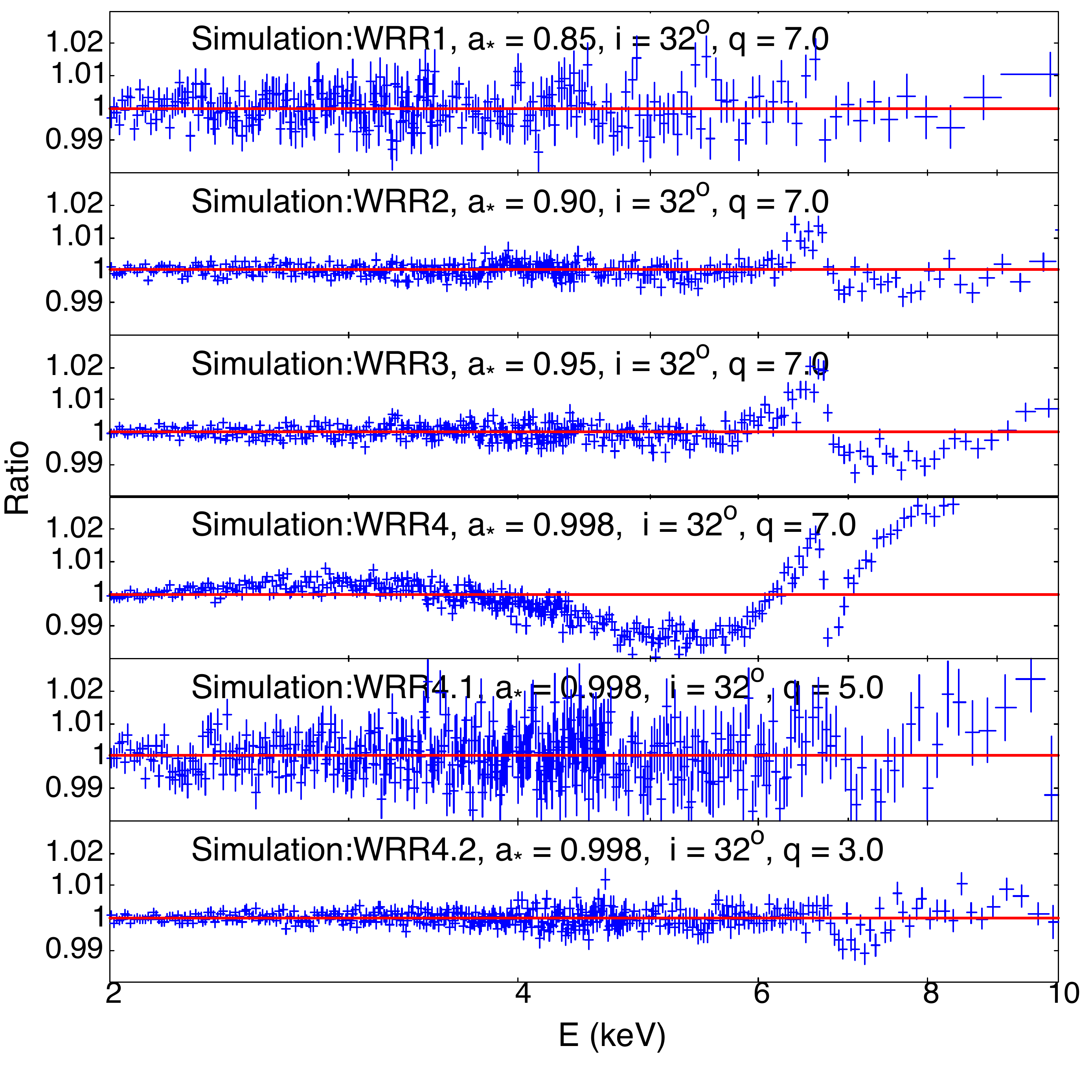}
\hspace{-0.2cm}
%\vspace{0.0cm}
\includegraphics[type=pdf,ext=.pdf,read=.pdf,width=9cm,height=9cm]{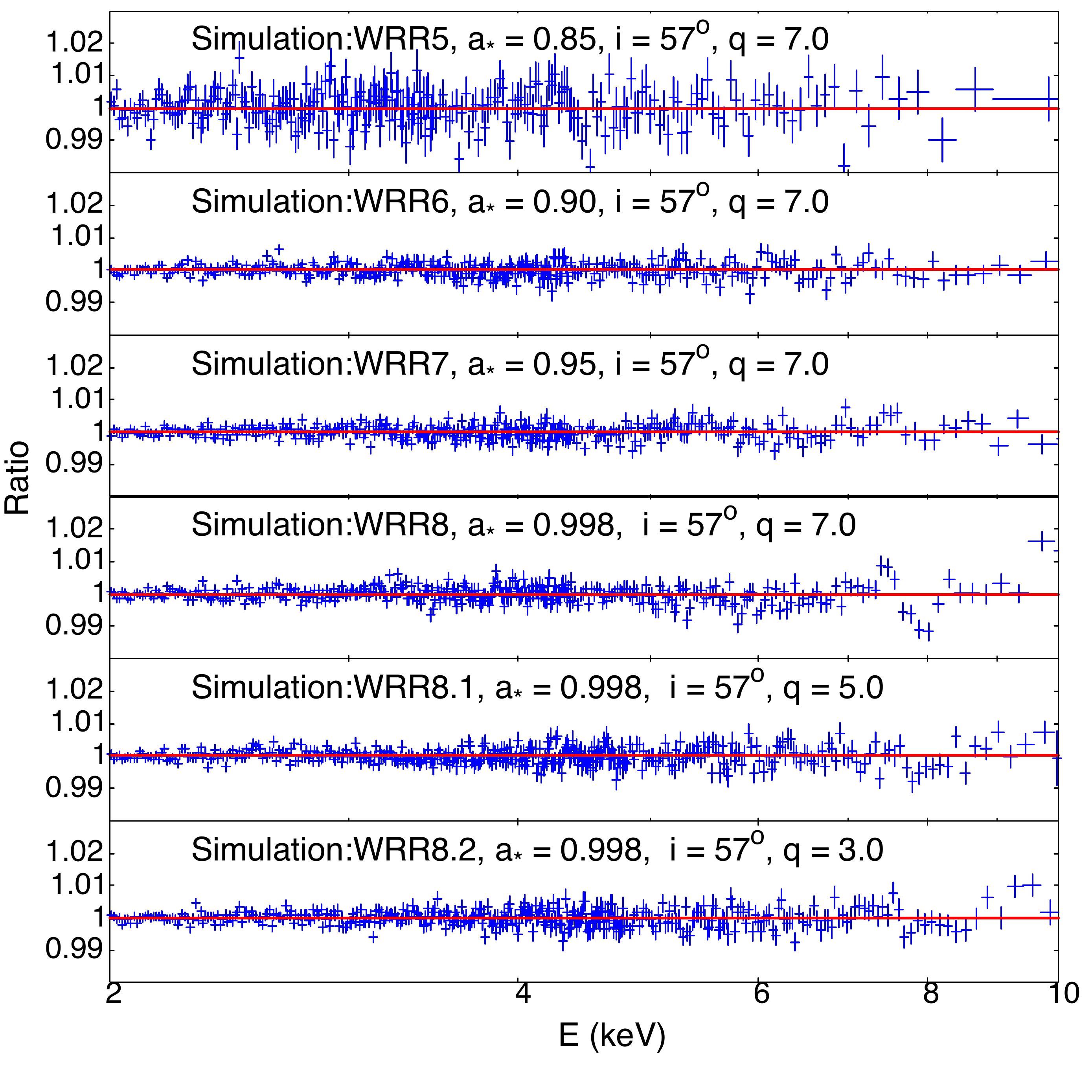}
\end{center}
\vspace{-0.1cm}
\caption{Data to best-fit model ratios for the simulations NRR1-NRR4,  {NRR4.1, NRR4.2} (top left panel), NRR5-NRR8, {NRR8.1, NRR8.2}  (top right panel), WRR1-WRR4, {WRR4.1, WRR4.2}  (bottom left panel), and WRR5-WRR8, {WRR8.1, WRR8.2}  (bottom right panel). } \label{f-ratio}
\end{figure*}

\begin{figure*}[t]
\begin{center}
\vspace{0.5cm}
\includegraphics[type=pdf,ext=.pdf,read=.pdf,width=9cm,height=9cm]{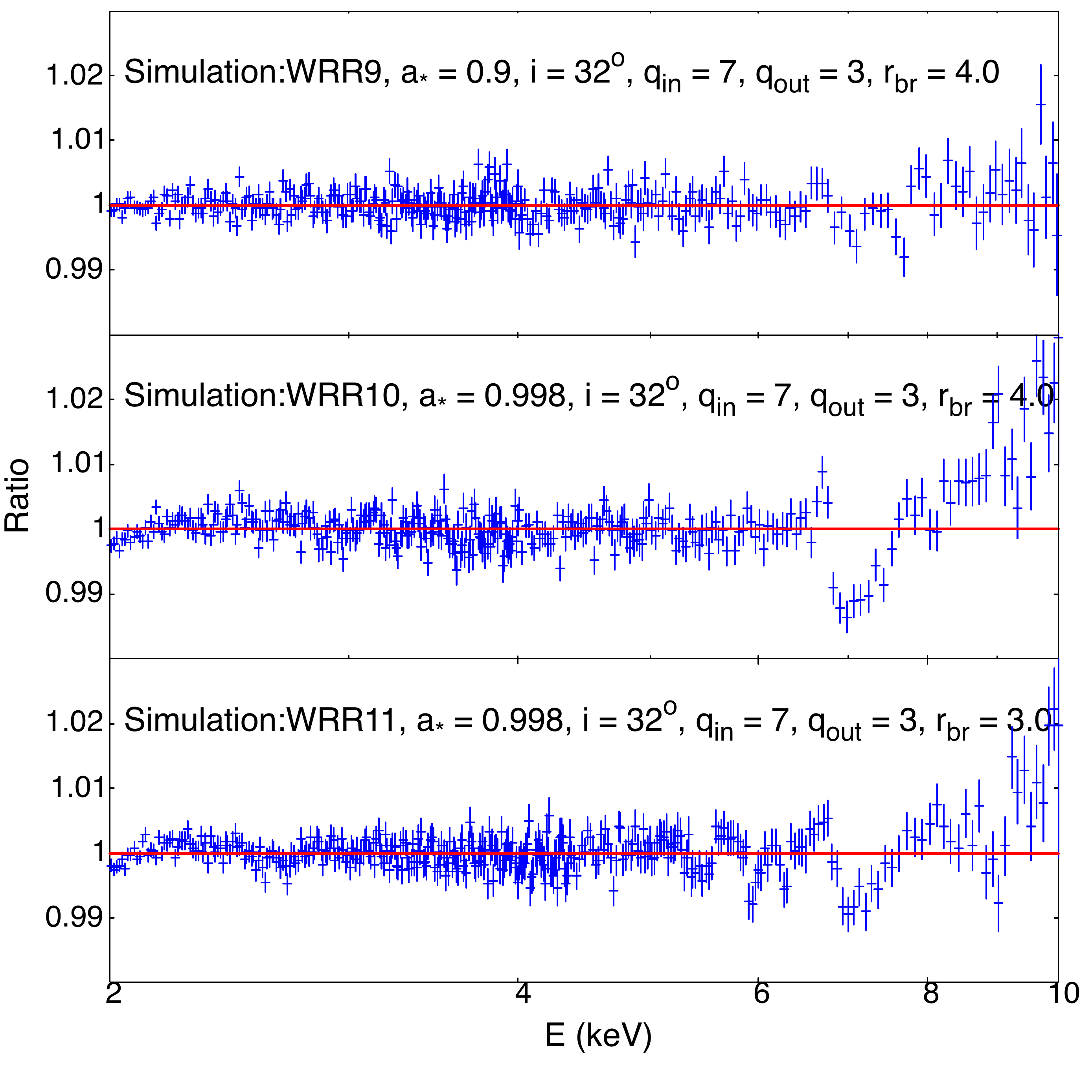}
\hspace{-0.2cm}
\vspace{-0.2cm}
\includegraphics[type=pdf,ext=.pdf,read=.pdf,width=9cm,height=9cm]{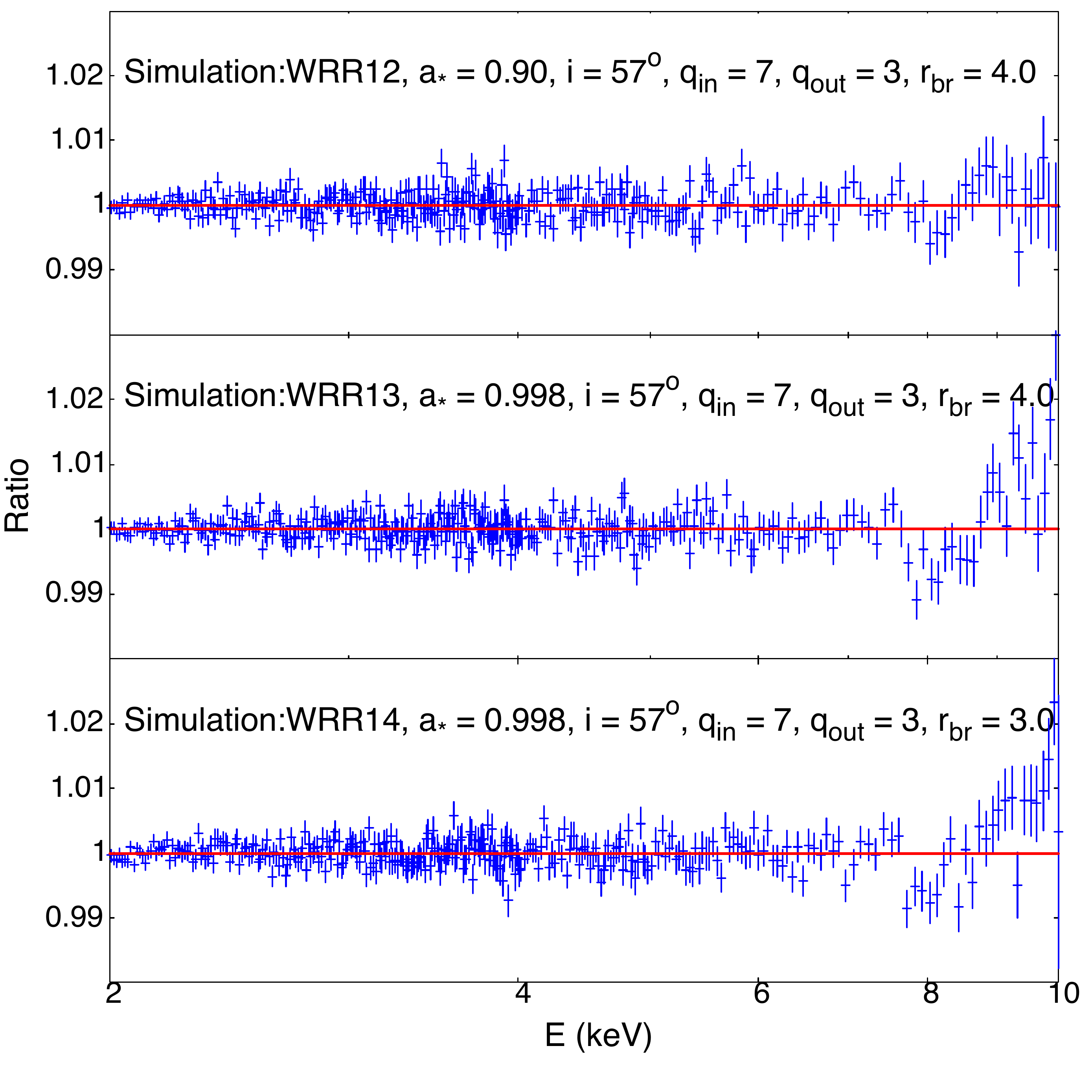}
\end{center}
\vspace{-0.1cm}
\caption{Data to best-fit model ratios for the simulations with a broken power-law emissivity profile  WRR9-WRR11 (left panel) and  WRR12-WRR14  (right panel). We omit the ratio plots for the simulations without returning radiation (NRR9-NRR14).  }
\label{f-ratio-rbr}
\end{figure*}

The results of our spectral analysis are reported in Tab.~\ref{t-fit-c} (simulations~NRR1-NRR4), Tab.~\ref{t-fit-c_min} (simulations NRR4.1, NRR4.2), Tab.~\ref{t-fit-d} (simulations~NRR5-NRR8), {Tab.~\ref{t-fit-d_min} (simulations NRR8.1, NRR8.2)}, Tab.~\ref{t-fit-wrr32} (simulations~WRR1-WRR4),  {Tab.~\ref{t-fit-wrr32_min} (simulations WRR4.1, WRR4.2)}, Tab.~\ref{t-fit-b} (simulations~WRR5-WRR8), and  {Tab.~\ref{t-fit-b_min} (simulations WRR8.1, WRR8.2)} for the simulations with input emissivity profile described by a single power-law. For the simulations with the input emissivity profile described by a broken power-law, the results are shown in Tab.~\ref{t-fit-rbr-i32} (WRR9-WRR11) and Tab.~\ref{t-fit-rbr-i57} (WRR12-WRR14).  We do not report the best-fit tables for the simulations excluding the returning radiation with a broken power-law setup (NRR9-NRR14), but we performed the analysis and always recovered the input parameters within the 90$\%$ confidence interval. The impact of the returning radiation on the estimate of the black hole spin parameter $a_*$, inclination angle of the disk $i$, emissivity index $q$, and iron abundance of the disk $A_{\rm Fe}$ is summarized in Fig.~\ref{f-spins}, Fig.~\ref{f-angle}, Fig.~\ref{f-em-index}, and Fig.~\ref{f-afe}, respectively. In all plots, the red circles show the estimate of the parameters in the simulations that include the returning radiation and the green squares indicate the measurements from the simulations that ignore the returning radiation. In the rightmost panel of Figs.~\ref{f-spins}-\ref{f-afe}, for a broken power-law setup, the red circle represents the simulations with breaking radius $r_{\rm br} = 4~R_{\rm g}$ and the black circle represents the simulations with breaking radius $r_{\rm br} = 3~R_{\rm g}$; we omit to show the best-fit parameters from simulations without returning radiation.  The thick dashed lines correspond to the case in which we would perfectly recover the correct input parameter. Fig.~\ref{f-ratio} shows the data to best-fit model ratios of the 24~fits for a single power-law emissivity profile case, and we can see that some fits are good and other fits are not. Fig.~\ref{f-ratio-rbr} shows the data to best-fit model ratios for the simulations that include the returning radiation with a broken power-law emissivity profile case. We discuss these results in the following section.

%%%%%%%%%%%%%%%%%%%% spins %%%%%%%%%%%%%%%%%%%
%%%%%%%%%%%%%%%%%%%%%%%%%%%%%%%%%%%%%%%%%%

%%%%%%%%%%%%%%%%%%%% emissvity index %%%%%%%%%%%%%%%%%%%
%%%%%%%%%%%%%%%%%%%%%%%%%%%%%%%%%%%%%%%%%%

%%%%%%%%%%%%%%%%%%%%     ref_frac    %%%%%%%%%%%%%%%
%\begin{figure*}[t]
%\vspace{0.5cm}
%\begin{center}
%\includegraphics[type=pdf,ext=.pdf,read=.pdf,width=8cm]{ref_frac_i32_q7}
%\hspace{0.5cm}
%\includegraphics[type=pdf,ext=.pdf,read=.pdf,width=8cm]{ref_frac_i57_q7}
%\end{center}
%\vspace{-0.4cm}
%\caption{ Input spin vs output reflection fraction. Red filled circles represent the results with returning radiations while green square without returning radiations returning radiations. \label{f-ref-frac}}
%\end{figure*}
%%%%%%%%%%%%%%%%%%%%%%%%%%%%%%%%%%%%%%%%%%

%%%%%%%%%%%%%%%%% Ratio plots %%%%%%%%%%%%%%%%%%%

\begin{figure*}
\begin{center}
\includegraphics[type=pdf,ext=.pdf,read=.pdf,width=9cm]{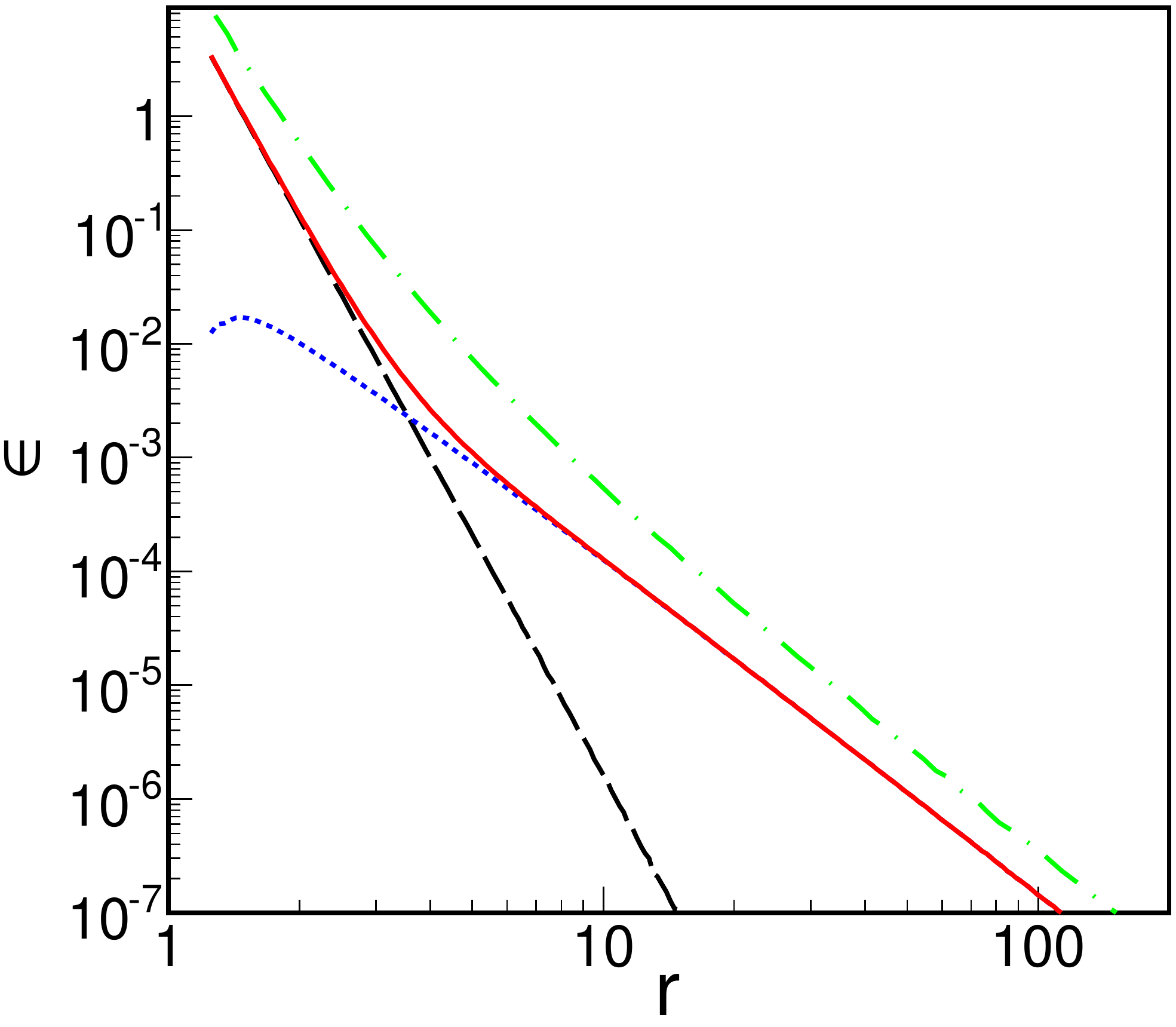}
\end{center}
%\vspace{-0.1cm}
\caption{Radial emissivity profiles for simulation WRR4. The dashed black line shows the input emissivity profile $q=7$ for the irradiation from the corona. The dotted blue curve is the emissivity profile for the irradiation of the returning radiation. The solid red curve is their sum, corresponding thus to the actual emissivity profile when we include the returning radiation. The dot-dashed green curve shows the emissivity profile for the lamppost model with $h=1.6$~$R_{\rm g}$. } \label{radial_profile}
\end{figure*}

\begin{figure*}[t]
\begin{center}
\includegraphics[type=pdf,ext=.pdf,read=.pdf,width=8.5cm]{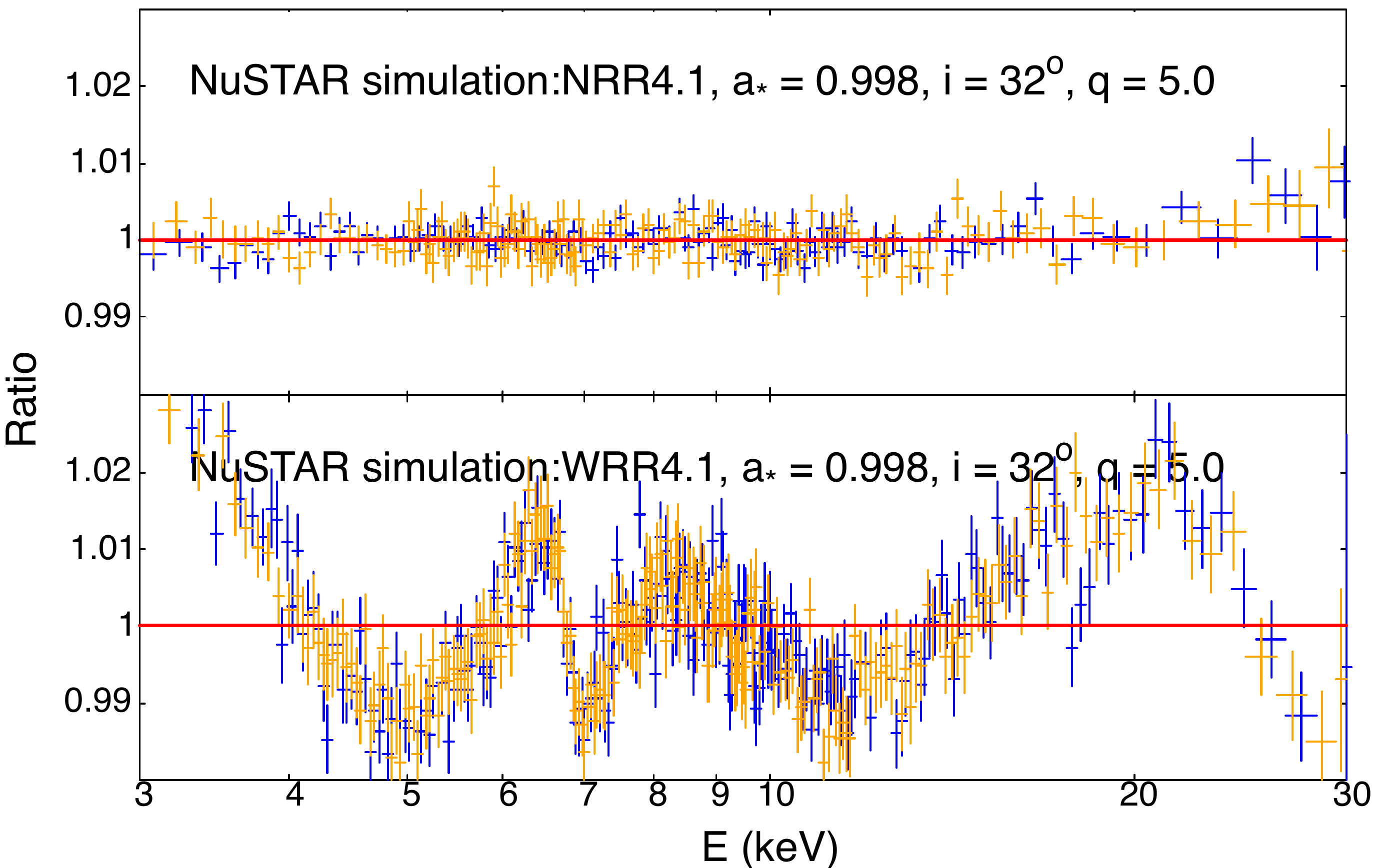}
\hspace{0.2cm}
\includegraphics[type=pdf,ext=.pdf,read=.pdf,width=8.5cm]{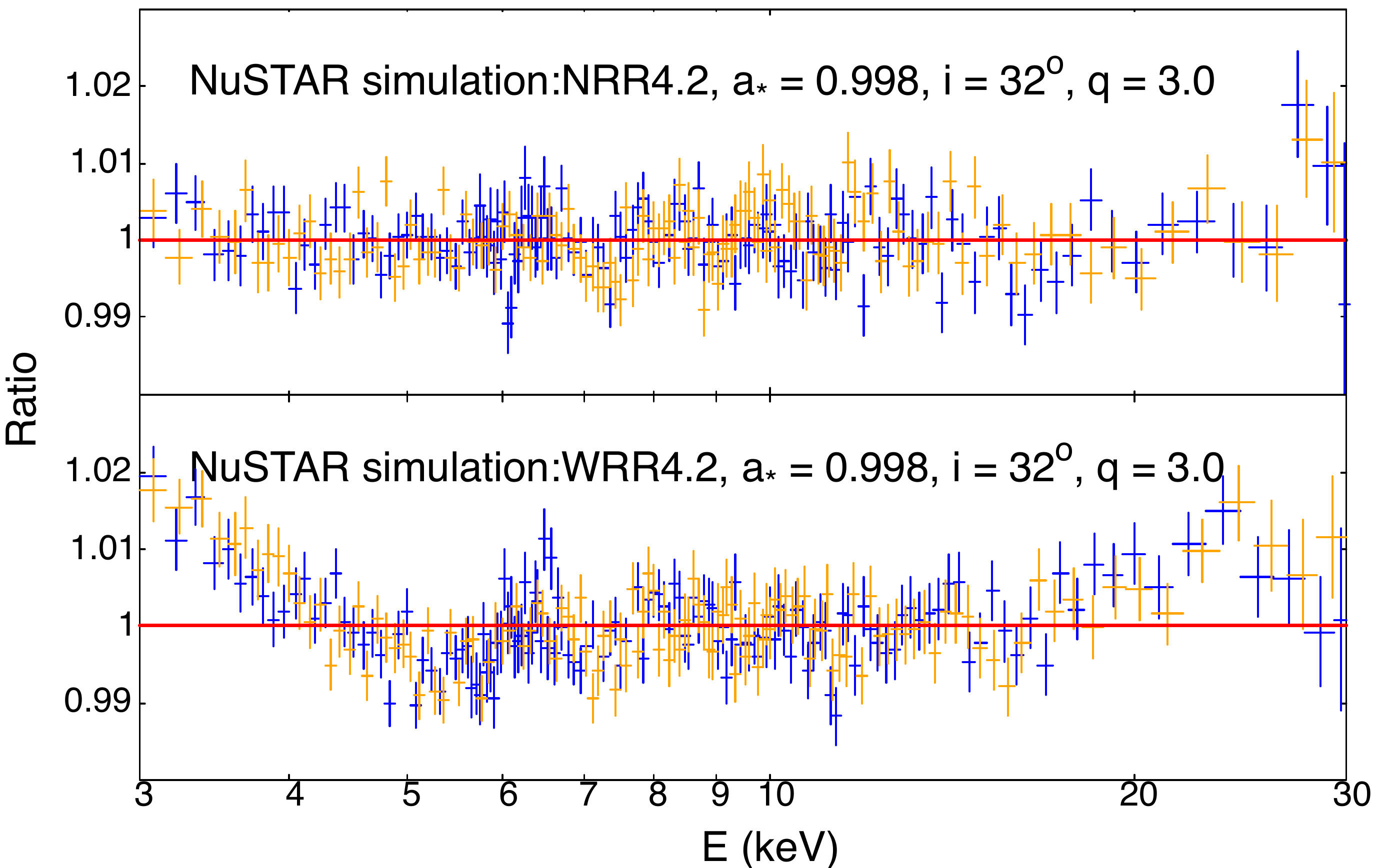}
\end{center}
\vspace{-0.3cm}
\caption{Data to best-fit model ratios for simulated observations with \textsl{NuSTAR} using the input parameter of NRR4.1, NRR4.2, WRR4.1, and WRR4.2. Note that the exposure time is only 10~ks.  \label{ns-ratio}}
\end{figure*}

%%%%%%%%%%%%%%%%%%%%%%%%%%%%%%%%%%%%%%%%%%%%

\section{Discussion}\label{s-con}

\subsection{Quality of the fits}

First of all, we can evaluate the quality of the fits from the reduced $\chi^2$ of every best-fit and from the ratio plots in Fig.~\ref{f-ratio} and Fig.~\ref{f-ratio-rbr}.

For the simulations without returning radiation (NRR1-NRR8, NRR4.1, NRR4.2, NRR8.1, NRR8.2), the quality of the fit is good as expected because we are employing the correct model to fit the simulated observations. We do not always recover the correct input parameter, but the difference between the fitted and the actual values does not exceed 10\%, see Tabs.~\ref{t-fit-c}-\ref{t-fit-d_min}. This gives the magnitude of bias introduced by the \texttt{fakeit} procedure, in which, e.g., the Poisson noise is added to synthetic spectra.

For the simulations with returning radiation, we need to discuss separately those with $q=7$ and those with lower value of the emissivity index or broken power-law. For the simulations with $q = 7$, we clearly see that the quality of the fit is worse when $i = 32^\circ$ and better for $i = 57^\circ$. As we increase the input value of the black hole spin parameter, the quality of the fits get worse. For example, in WRR5 the input spin is $a_* = 0.85$ and the input inclination angle is $i = 57^\circ$. The reduced $\chi^2$ is around 1 and the ratio plot does not show strong residuals. The fit can recover well all input parameters. Our results for WRR4, with the input parameters $a_* = 0.998$ and $i = 32^\circ$, are qualitatively different. The reduced $\chi^2$ is around 6 and the ratio plot shows a very pronounced residual pattern. Note, however, that we are considering very high quality data with 100~million photons. It is likely that residuals with this amplitude, $\la 3$\%, would not be revealed in currently available observations. Interestingly, though, similar looking residual patterns are found in the long ($\sim 300$~ks) \textsl{XMM} datasets of the Seyfert galaxy 1H 0707--495 by \citet{2020arXiv200615016S}, see their Fig.~4: in both cases we see an excess of counts at 6-7~keV and at 8-10~keV. The distortion introduced by the returning radiation in WRR4 cannot be compensated by a different set of the values of the model parameters. Some input parameters are recovered well but other input parameters are definitively biased. For example, the best-fit value of the iron abundance is pegged to the minimum allowed by the model, 0.5. The emissivity profile is highly underestimated.

For WRR4, we find a better fit, with $\chi_{\nu}^2 \simeq 2$, using a broken power-law emissivity profile. For the inner and outer emissivity indices, we find $q_{\rm in}  = 6.86^{+0.03}_{-0.16}$ and $q_{\rm out} = 2.88^{+0.01}_{-0.02}$, respectively, and for the breaking radius we get $r_{\rm br} = 3.97^{+0.02}_{-0.03}~R_{\rm g}$. The other parameters are rather weakly affected (i.e.\ have values similar to those in Tab.~\ref{t-fit-wrr32} for WRR4), except the reflection strength, which increases to $\mathcal{R} \simeq 2.2$. Fig.~\ref{radial_profile} shows the emissivity profiles for this simulation. The dashed black line shows the input emissivity profile, $q=7$, related to the illumination of the accretion disk by the corona. The dotted blue curve is the irradiation of the returning radiation -- at large radii it can be described with an emissivity index $q \simeq 3$. The red curve is their sum, i.e. the actual simulated radial profile for WRR4. We see that the fitted parameters of the broken power-law are very close to the actual profile. We return to this issue below. We may also compare the fitted profile to profiles predicted by the lamppost geometry, often regarded as the main motivation for steep radial profiles which flatten in the outer parts. The dot-dashed green curve is for the lamppost model with $h=1.6$~$R_{\rm g}$, which produces the radial distribution closest to that fitted in WRR4. As we see, the broken-power law gives only a very crude approximation for the lamppost profile, in which we rather have a continuous change of the emissivity index. This is in contrast to the returning radiation profile, which is much  closer to the broken power-law.

If we consider the simulations with $q=5$ and 3, we see that decreasing the input value of the emissivity profile reduces the effect of the returning radiation: the quality of the fits improves. This is perfectly understandable with the behavior of the spin parameter $a_*$. For high values of $a_*$ and $q$, a significant fraction of radiation is emitted very close to the black hole, where light bending is strong and thus a significant fraction of the radiation returns to the disk. The distortion of the spectrum from the returning radiation is important. For lower values of the black hole spin parameter and the emissivity index $q$, the fraction of returning radiation is reduced, so its distortion to the reflection spectrum is moderate.

\subsection{Measurement of the properties of the system}

The impact of the returning radiation on the estimate of the black hole spin is illustrated in Fig.~\ref{f-spins}. The larger panels show the fits of the simulations with input emissivity index $q = 7$. The green squares representing the simulated data without returning radiation follow the thick dashed diagonal line; that is, the model \texttt{reflkerr} without returning radiation can recover relatively well the correct spin when the returning radiation is turned off in the simulated data as well. The red circles are instead the results of the fit of the simulated observations in which the returning radiation is turned on, and we can see a difference between the low ($i = 32^\circ$) and high ($i = 57^\circ$) viewing angle. For the simulations with $i = 57^\circ$, we recover the correct input value, so the impact of the returning radiation on the spin measurement is negligible. For the simulations with $i = 32^\circ$, it seems that the returning radiation, when it is not taken into account in the fitting model, leads to overestimating the black hole spin parameter except when the input spin parameter is $a_* = 0.998$, where we find the opposite case and we slightly underestimate the black hole spin. The bias is not as large as we may have from the structure of the accretion disk~\citep{2019arXiv191106605R,2020MNRAS.491..417R}, but it can be important in the case of very precise spin measurements. The cases with $q=5$ and 3 (smaller panels) do not show any remarkable property concerning the estimate of the black hole spin parameter.

The capability of the fits to recover the inclination angle is shown in Fig.~\ref{f-angle}. For the simulations with $i = 57^\circ$, we do not see a clear bias in the estimate of the inclination angle from the returning radiation. Note that the input value of the inclination angle is not recovered very well, even when the simulation is without returning radiation. For the simulations with $i = 32^\circ$, we find that the returning radiation leads to an overestimate of the inclination angle $i$.

The impact of the returning radiation on the estimate of the emissivity index $q$ is shown in Fig.~\ref{f-em-index}. For the simulated observations without returning radiation (green squares), we surely recover well the input value of $q$. For the simulated data with returning radiation (red circles), the fits of the simulations with $q=7$ and 5 do not recover well the input emissivity profile, and this is true for both inclination angles. We note that we significantly underestimate the value of the emissivity index $q$ only for the simulations with $a_* = 0.998$. As we noted above, however, a much better fit for this case is found for a broken power-law emissivity, in which the inner index equals the actual input value. Then, if the radial redistribution of reflection by returning radiation is important, it leads to a flattening of the emissivity beyond a few gravitational radii. Interestingly, such a value of the breaking radius has been often reported in models assuming a broken power-law radial emissivity \citep[e.g.][]{2013ApJ...775L..45M,2015ApJ...806..149K,2016A&A...589A..14D}. When $q=3$, we are able to recover the correct input value of $q$ and this is true for both inclination angles.

Lastly, Fig.~\ref{f-afe} shows the results of the iron abundance $A_{\rm Fe}$. For $q=7$, we clearly see that the iron abundance is significantly underestimated when $a_* = 0.998$, and this is true for both $i = 32^\circ$ and $i = 57^\circ$. If we decrease the input emissivity profile, the bias decreases and for $q=3$ the estimate of $A_{\rm Fe}$ is not too bad. We do not find that the returning radiation leads to an overestimate of the iron abundance in a manner proposed by \citet{2002MNRAS.336..315R}. As we noted above, their claim was based on a qualitative resemblance of multiple-reflection and overabundant reflection spectra. Our model ratio plots in Fig. \ref{f-ratio} show that we do reproduce qualitative properties similar to those estimated by \citet{2002MNRAS.336..315R}. Namely, we see positive residuals at the position of the Fe K$\alpha$ emission line and negative residuals at the position of the Fe K$\alpha$ absorption edge. These residuals are stronger when the returning radiation is more important, as could be expected. However, we find that the residuals cannot be reduced by increasing $A_{\rm Fe}$. Increasing Fe abundance and the increasing contribution of returning radiation produce very different spectral distortions especially at $E \ga 7.5$ keV, where the former reduces while the latter enhances the low energy part of the Compton hump. To compensate for this enhancement, low values of iron abundance,  $A_{\rm Fe} < 1$, and high reflection fraction, $\mathcal{R} > 1$, are fitted in the standard (i.e.\ without returning radiation) reflection model. This effect is extreme for $a_* = 0.998$, where the fitted iron abundance pegged at the lowest allowed value, i.e.\ 0.5. We note, however, that this property should strongly depend on data quality. It is possible that for a very poor quality, the returning radiation could mimic the high Fe abundance to some extent, especially that current X-ray detectors often have reduced performance in the 8--10 keV range, which is crucial for distinguishing between the two effects discussed here.

While it is a longstanding issue that the analysis of the reflection spectrum of many sources, either black hole binaries and AGNs, finds unexpectedly high values of $A_{\rm Fe}$~\citep{2018ASPC..515..282G}, we also have a few opposite cases. For GRS~1915+105~\citep{2019ApJ...884..147Z} and GS~1354--645~\citep{2016ApJ...826L..12E,2018ApJ...865..134X}, the analysis of the reflection spectrum has provided a spin parameter very close to 1, a very high viewing angle, $i \sim 75^\circ$, and quite a low iron abundance, $A_{\rm Fe} \sim 0.6$. From Fig.~\ref{f-afe}, we may argue that the low iron abundance found in GRS~1915+105 and in GS~1354--645 might be due to the returning radiation. The fact that the inclination angle is very high in both sources may lead to a good fit, even if we do not include the returning radiation. Sources for which high values of $A_{\rm Fe}$ are found do not have such high values of $i$. Once again, we recall that our simulations and fitting model are for neutral disks, and we would need to study the case of ionized disks in order to better support this conjecture. 

Now, we move to the simulations that are performed with input emissivity profile described by a broken power-law. The ratio plots for these simulations, including the returning radiation (WRR9-WRR14), are shown in Fig.~\ref{f-ratio-rbr}. As in the case of a single power-law emissivity profile, the quality of the fit is worse when $ i = 32^{\circ}$ and better for $i = 57^{\circ}$. For example, the reduced $\chi^2$ is around~1.47 for the case ($a_*$ = 0.998, $i = 32^{\circ}$, and $r_{\rm br} = 4~R_{\rm g}$) and is around~1.08 for the case ($a_* = 0.998$, $i = 57^{\circ}$, and $r_{\rm br} = 4~R_{\rm g}$). As we increase the input value of the spin parameter, the quality of the fits becomes poor. Moreover, as we reduce the value of input breaking radius, the quality of the fits improves. This is because the emission from the inner part of the disk is reduced, which decreases the fraction of the returning radiation in the total spectrum. The ratio plots in the simulations with $a_* = 0.998$ show evident residuals near the iron line region, which means the model without returning radiation cannot compensate for the distortion in the reflection spectrum produced by the returning radiation. We also recall that we are considering very high-quality data with 100 million photons, and such residuals would not likely be visible in currently available data.

The capability of fits to recover the input parameters for the broken power-law case are shown in Figs.~\ref{f-spins}-\ref{f-afe} in the rightmost panels. The red circle represents the results from simulations with the breaking radius $r_{\rm br} = 4~R_{\rm g}$, and the black circle represents the result from the simulations with the breaking radius $r_{\rm br} = 3~R_{\rm g}$.  The fit can recover the black hole spin parameter and the inclination angle in all these simulations. When we compare ($a_* = 0.9$, $i = 32^{\circ}$, $r_{\rm br} = 4~R_{\rm g}$) with the case ($a_* = 0.9$, $i = 32^{\circ}$, $q = 7$), we see that the single power-law emissivity profile overestimates the black hole spin. The same is also true for the recovered inclination angle. For most cases, the inner and the outer emissivity indices are recovered except for the case ($a_* = 0.998$, $i = 57^{\circ}$, $r_{\rm br} = 4~R_{\rm g}$) where the inner emissivity index is slightly underestimated; the same trend is also observed in the case ($a_* = 0.998$, $i = 57^{\circ}$, $q = 7$). Finally, the iron abundance is always underestimated and the deviation increases with increasing input spin parameter. We also notice the same trend for the simulations with a single power-law emissivity profile.

\subsection{Simulations with \textsl{NuSTAR}}

We note that current (e.g., \textsl{XMM-Newton} or \textsl{NICER}) and future (e.g., \textsl{Athena}) X-ray missions are often limited to the energy band below $\sim$10~keV. The results reported so far, which are valid when we fit data up to 10~keV, are thus relevant for all these instruments. However, we can briefly investigate how our conclusions would change if we have data above 10~keV. To address this issue, we simulated some observations with \textsl{NuSTAR}. We employ the input parameters of $i=32^\circ$, $a_* = 0.998$, $q=5$ and $q=3$, for which the contribution of the returning radiation is moderate. The exposure time is set to be 10~ks and we set the photon flux to $4 \cdot 10^{-10}$~erg~cm$^{-2}$~s$^{-1}$ in the energy range 0.2-12~keV, which leads to have 10 million counts in the 3--30~keV band. We fit the data up to 30~keV. The discrepancy between the reflection produced by the power-law and by returning radiation is so strong above 30 keV that no adequate description of the data in the full \textsl{NuSTAR} energy range of 3--79 keV could be found\footnote{{We checked that the discrepancy can be partially reduced if the reflection of thermal Comptonization spectrum, {\tt compps}, is used instead of the power-law reflection, but this is not the input model.}}. The best-fit values of the parameters are reported in Tab.~\ref{t-fit-ns}. Data to best-fit model ratio plots are shown in Fig.~\ref{ns-ratio}. The quality of the fit for the simulations without returning radiation is good and the input parameters are recovered well. The quality of the fit for WRR4.1 (with $q=5$) is poor, with the reduced $\chi^{2}$ of around 2.3. Then, we found a much better fit using a broken power-law emissivity, see Tab.~\ref{ns-bpl}. However, in contrast to the \textsl{NICER} fit discussed above, the estimates of the emissivity indices are now quite different from the actual radial distribution in WRR4.1. Moreover, the other fitted parameters in WRR4.1 are very different from the input values. For example, the iron abundance is fitted at $A_{\rm Fe} \simeq 3$ with a single power-law and at $A_{\rm Fe} \simeq 9$ with a broken power-law model.

\subsection{Summary and conclusions}

To conclude, we first restate that the returning radiation distorts the relativistic reflection spectra through the combination of (i) radial redistribution of the irradiating flux, and (ii) spectral distortion of the incident radiation. Our results with the \textsl{NICER} simulations indicate that the effect (ii) is rather modest if only data below 10 keV are considered, whereas the effect (i) is dominant in the most extreme cases. In these cases, the actual radial (re)distribution can be closely recovered if the broken power-law emissivity is applied. This conclusion seems to be general, because we considered here a neutral disk for which effect (ii) is maximized. However, if the fitted energy range extends above 10 keV, the effect (ii) outweights the other effects and the actual radial distribution as well as the crucial parameters, like $A_{\rm Fe}$ or $\Gamma$, are not recovered. We can expect a strong dependence of effect (ii) on the disk ionization state, therefore, a detailed analysis in this regime would require a model completed by ionization effects.

Lastly, we note that some of our simulations with returning radiation lead to extremely poor fits, which can question about how realistic our model with neutral disk is because such bad fits are not found when real observations are analyzed. However, there are a few differences between our simulations and real observations. For the \textsl{NICER} simulations, we considered 300~ks observations, while normal \textsl{NICER} data are 10-30~ks. The exceptionally high photon count leads to reduced $\chi^2$ with values like 6. Considering a 30~ks observation, the fit would not be too bad. Moreover, when we analyze a real observation we try to improve the fit by adjusting the model or by adding extra components. As we have seen, employing a broken power-law emissivity profile in the simulations with returning radiation can help to get a better fit. In this study, we have focused on the actual capability of a reflection model without returning radiation to recover the correct input parameters when returning radiation is included. A closer comparison with real observations, which is beyond the scope of the current work, would have required a blind analysis~\citep[see, e.g.,][]{2018A&A...614A..44K}.

%%%%%%%%%%%%%%%%%%%%%%%%%%%%%%%

\vspace{0.5cm}

{\bf Acknowledgments --}
This work was supported by the Innovation Program of the Shanghai Municipal Education Commission, Grant No.~2019-01-07-00-07-E00035, and the National Natural Science Foundation of China (NSFC), Grant No.~11973019. M.S. and A.N. thank the Polish National Science Centre Grants 2015/18/A/ST9/00746 and 2016/21/B/ST9/02388. A.N. and C.B. are members of the International Team~458 at the International Space Science Institute (ISSI), Bern, Switzerland, and acknowledge support from ISSI during the meetings in Bern.

\newpage

%%%%%%%%%%%%%%%%%%%%%%%%%%%%%%%

\newpage

%%%%%%%% NRR1-NRR4, i = 32, q = 7.0 %%%%%%%%%%%%%%%%%%%

\begin{table*}
\centering
\caption{ \label{t-fit-c}}
{\renewcommand{\arraystretch}{1.3}
\begin{tabular}{lcc|ccc|ccc|ccc}
\hline\hline
Simulation & \multicolumn{2}{c}{NRR1} && \multicolumn{2}{c}{NRR2} && \multicolumn{2}{c}{NRR3} && \multicolumn{2}{c}{NRR4} \\
 & Input & Fit && Input & Fit && Input & Fit && Input & Fit \\
\hline
\texttt{tbabs} &&&&&&&&&& \\
$N_{\rm H} / 10^{20}$ cm$^{-2}$ & $6.74$ & $6.74^\star$ && $6.74$ & $6.74^\star$ && $6.74$ & $6.74^\star$ && $6.74$ & $6.74^\star$ \\
\hline
\texttt{reflkerr} &&&&&&&& \\
$q$ & $7$ & $6.6^{+0.3}_{-0.3}$ && $7$ & $7.16^{+0.11}_{-0.05}$ && $7$ & $6.72^{+0.07}_{-0.08}$ && $7$ & $7.03^{+0.21}_{-0.21}$ \\
$i$ [deg] & $32$ & $29.7^{+0.7}_{-0.6}$ && $32$ & $33.50^{+0.22}_{-0.16}$ && $32$ & $30.32^{+0.22}_{-0.20}$ && $32$ & $32.1^{+1.8}_{-1.5}$ \\
$a_*$ & $0.85$ & $0.829^{+0.007}_{-0.007}$ && $0.90$ & $0.9094^{+0.0011}_{-0.0011}$ && $0.95$ & $0.9455^{+0.0008}_{-0.0009}$ && $0.998$ & $0.9956^{+0.0023}_{-0.0027}$ \\
$A_{\rm Fe}$ & $1$ & $0.94^{+0.04}_{-0.03}$ && $1$ & $0.993^{+0.010}_{-0.009}$ && $1$ & $0.988^{+0.013}_{-0.010}$ && $1$ & $1.02^{+0.05}_{-0.08}$ \\
$\Gamma$ & $2$ & $2.015^{+0.008}_{-0.013}$ && $2$ & $2.000^{+0.003}_{-0.006}$ && $2$ & $2.000^{+0.004}_{-0.004}$ && $2$ & $1.965^{+0.024}_{-0.008}$ \\
%$E_{\rm cut}$ [MeV] & $100$ & $100^{\star}$ && $100$ & $100^{\star}$ && $100$ & $100^{\star}$ && $100$ & $100^{\star}$ \\
$\mathcal{R}$ & $1$ & $1.07^{+0.05}_{-0.08}$ && 1 & $1.000^{+0.015}_{-0.035}$ && 1 & $0.985^{+0.030}_{-0.021}$ && 1 & $0.78^{+0.15}_{-0.10}$ \\
\hline
$\chi^2/\nu$ && $\quad 1096.66/992 \quad$ &&& $\quad 941.39/992 \quad$ &&& $\quad 1024.57/992 \quad$ &&& $\quad 1007.56/992 \quad$ \\
&& =1.10551 &&& =0.94898 &&& =1.03284 &&& =1.01569 \\
\hline\hline
\vspace{0.2cm}
\tablenotetext{0}{Best-fit values for simulations~NRR1-NRR4 (returning radiation is not included). The inclination angle of the disk is $i = 32^{\circ}$. The reported uncertainties correspond to 90\% confidence level for one relevant parameter. $^\star$ indicates that the parameter is frozen in the fit. }
\end{tabular}}
\end{table*}

%%%%%%%% a = 0.998, i = 32, q = 5 and q = 3  NRR4.1, NRR4.2 %%%%%%%%
\begin{table*}
\centering
\caption{ \label{t-fit-c_min}}
{\renewcommand{\arraystretch}{1.3}
\begin{tabular}{lcc|ccc}
\hline\hline
Simulation & \multicolumn{2}{c}{NRR4.1} && \multicolumn{2}{c}{NRR4.2} \\
 & Input & Fit && Input & Fit \\
\hline
\texttt{tbabs} &&&& \\
$N_{\rm H} / 10^{20}$ cm$^{-2}$ & $6.74$ & $6.74^\star$ && $6.74$ & $6.74^\star$  \\
\hline
\texttt{reflkerr} &&&& \\
$q$ & $5$ & $5.10^{+0.08}_{-0.10}$ && $3$ & $3.025^{+0.018}_{-0.016}$ \\
$i$ [deg] & $32$ & $32.3^{+2.1}_{-2.5}$ && $32$ & $31.99^{+0.21}_{-0.17}$ \\
$a_*$ & $0.998$ & $0.997^{+ (\rm P)}_{-0.005}$ && $0.998$ & $0.980^{+ (\rm P)}_{-0.018}$  \\
$A_{\rm Fe}$ & $1$ & $1.03^{+0.03}_{-0.05}$ && $1$ & $1.02^{+0.010}_{-0.009}$ \\
$\Gamma$ & $2$ & $2.008^{+0.009}_{-0.012}$ && $2$ & $1.9997^{+0.0029}_{-0.0014}$ \\
%$E_{\rm cut}$ [MeV] & $100$ & $100^{\star}$ && $100$ & $100^{\star}$ && $100$ & $100^{\star}$ && $100$ & $100^{\star}$ \\
$\mathcal{R}$ & $1$ & $1.06^{+0.05}_{-0.07}$ && $1$ & $0.997^{+0.018}_{-0.011}$ \\
\hline
$\chi^2/\nu$ && $\quad 702.35/792 \quad$ &&& $\quad 760.48/792 \quad$ \\
&& =0.88681 &&& =0.96020 \\
\hline\hline
\vspace{0.2cm}
\tablenotetext{0}{{Best-fit values for simulations~NRR4.1 and NRR4.2 (returning radiation is not included). The inclination angle of the disk is $i = 32^{\circ}$. The reported uncertainties correspond to 90\% confidence level for one relevant parameter. $^\star$ indicates that the parameter is frozen in the fit. $(\rm P)$ indicates that the 90\% confidence level reaches the boundary of the parameter range (the spin parameter $a_*$ is allowed to vary in the range $-0.998$ to 0.998). }}
\end{tabular}}
\end{table*}

%%%%%%%% Simulations set: NRR5-NRR8 i = 32, q = 7 %%%%%%%%%%%%%%%%%%%
\begin{table*}
\centering
\caption{ \label{t-fit-d}}
{\renewcommand{\arraystretch}{1.3}
\begin{tabular}{lcc|ccc|ccc|ccc}
\hline\hline
Simulation & \multicolumn{2}{c}{NRR5} && \multicolumn{2}{c}{NRR6} && \multicolumn{2}{c}{NRR7} && \multicolumn{2}{c}{NRR8} \\
 & Input & Fit && Input & Fit && Input & Fit && Input & Fit \\
\hline
\texttt{tbabs} &&&&&&&&&& \\
$N_{\rm H} / 10^{20}$ cm$^{-2}$ & $6.74$ & $6.74^\star$ && $6.74$ & $6.74^\star$ && $6.74$ & $6.74^\star$ && $6.74$ & $6.74^\star$ \\
\hline
\texttt{reflkerr} &&&&&&&& \\
$q$ & $7$ & $6.8^{+0.7}_{-0.5}$ && $7$ & $7.45^{+0.16}_{-0.13}$ && $7$ & $6.83^{+0.06}_{-0.07}$ && $7$ & $6.47^{+0.12}_{-0.10}$ \\

$i$ [deg] & $57$ & $55.6^{+0.6}_{-0.4}$ && $57$ & $57.70^{+0.13}_{-0.13}$ && $57$ & $55.80^{+0.12}_{-0.11}$ && $57$ & $54.2^{+0.9}_{-1.4}$ \\

$a_*$ & $0.85$ & $0.827^{+0.006}_{-0.009}$ && $0.90$ & $0.9081^{+0.0014}_{-0.0021}$ && $0.95$ & $0.9454^{+0.0006}_{-0.0006}$ && $0.998$ & $0.998^{}_{-0.005}$ \\

$A_{\rm Fe}$ & $1$ & $1.04^{+0.10}_{-0.09}$ && $1$ & $1.005^{+0.030}_{-0.020}$ && $1$ & $0.992^{+0.020}_{-0.015}$ && $1$ & $0.931^{+0.045}_{-0.024}$ \\

$\Gamma$ & $2$ & $2.000^{+0.004}_{-0.005}$ && $2$ & $2.0000^{+0.0018}_{-0.0011}$ && $2$ & $2.0004^{+0.0025}_{-0.0006}$ && $2$ & $1.9917^{+0.0009}_{-0.0041}$ \\
%$E_{\rm cut}$ [MeV] & $100$ & $100^{\star}$ && $100$ & $100^{\star}$ && $100$ & $100^{\star}$ && $100$ & $100^{\star}$ \\
$\mathcal{R}$ & $1$ & $0.98^{+0.03}_{-0.04}$ && 1 & $0.983^{+0.014}_{-0.008}$ && 1 & $0.988^{+0.015}_{-0.004}$ && 1 & $0.89^{+0.12}_{-0.04}$ \\
\hline
$\chi^2/\nu$ && $\quad 983.50/992 \quad$ &&& $\quad 1020.64/992 \quad$ &&& $\quad  1007.81/992 \quad$ &&& $\quad 1006.00/992 \quad$ \\
&& =0.99143 &&& =1.02887 &&& =1.01594 &&& =1.01411 \\
\hline\hline
\vspace{0.2cm}
\tablenotetext{0}{{Best-fit values for simulations~NRR5-NRR8 (returning radiation is not included). The inclination angle of the disk is $i = 57^{\circ}$. The reported uncertainties correspond to 90\% confidence level for one relevant parameter. $^\star$ indicates that the parameter is frozen in the fit. The spin parameter $a_*$ is allowed to vary in the range $-0.998$ to 0.998. }}
\end{tabular}}
\end{table*}

%%%%%%%%%%%%%%% NRR8.1-NRR8.2 i = 57, q = 5, q = 3 %%%%%%%%%

\begin{table*}
\centering
\caption{ \label{t-fit-d_min}}
{\renewcommand{\arraystretch}{1.3}
\begin{tabular}{lcc|ccc}
\hline\hline
Simulation & \multicolumn{2}{c}{NRR8.1} && \multicolumn{2}{c}{NRR8.2}  \\
 & Input & Fit && Input & Fit  \\
\hline
\texttt{tbabs} &&&&\\
$N_{\rm H} / 10^{20}$ cm$^{-2}$ & $6.74$ & $6.74^\star$ && $6.74$ & $6.74^\star$  \\
\hline
\texttt{reflkerr} &&&&\\
$q$ &  $5$ & $5.01^{+0.13}_{-0.06}$ && $3$ & $3.02^{+0.11}_{-0.07}$ \\
$i$ [deg]& $57$ & $56.7^{+0.5}_{-0.4}$ && $57$ & $56.81^{+0.15}_{-0.14}$ \\
$a_*$ &  $0.998$ & $0.9953^{+0.0021}_{-0.0008}$ && $0.998$ & $0.989^{+ (\rm P)}_{-0.011}$ \\
$A_{\rm Fe}$ & $1$ & $0.980^{+0.038}_{-0.022}$ && $1$ & $0.99^{+0.05}_{-0.03}$\\
$\Gamma$ &  $2$ & $2.0009^{+0.0024}_{-0.0030}$ && $2$ & $1.998^{+0.004}_{-0.004}$ \\
%$E_{\rm cut}$ [MeV] & $100$ & $100^{\star}$ && $100$ & $100^{\star}$ && $100$ & $100^{\star}$ && $100$ & $100^{\star}$ \\
$\mathcal{R}$ &  $1$ & $1.001^{+0.013}_{-0.019}$ && 1 & $0.979^{+0.023}_{-0.028}$ \\
\hline
$\chi^2/\nu$ && $\quad 778.88/792 \quad$ &&& $\quad 797.26/792 \quad$ \\
&& =0.98343 &&& =1.0066 \\
\hline\hline
\vspace{0.2cm}
\tablenotetext{0}{{Best-fit values for simulations~NRR8.1 and NRR8.2 (returning radiation is not included). The inclination angle of the disk is $i = 57^{\circ}$. The reported uncertainties correspond to 90\% confidence level for one relevant parameter. $^\star$ indicates that the parameter is frozen in the fit. $(\rm P)$ indicates that the 90\% confidence level reaches the boundary of the parameter range (the spin parameter $a_*$ is allowed to vary in the range $-0.998$ to 0.998). }}
\end{tabular}}
\end{table*}

%%%%%%%% WRR1-WRR4, i = 32, q = 7 %%%%%%%%%%%%%%%%%%%

\begin{table*}
\centering
\caption{ \label{t-fit-wrr32}}
{\renewcommand{\arraystretch}{1.3}
\begin{tabular}{lcc|ccc|ccc|ccc}
\hline\hline
Simulation & \multicolumn{2}{c}{WRR1} && \multicolumn{2}{c}{WRR2} && \multicolumn{2}{c}{WRR3} && \multicolumn{2}{c}{WRR4}\\
 & Input & Fit && Input & Fit && Input & Fit && Input & Fit \\
\hline
\texttt{tbabs} &&&&&&&&&& \\
$N_{\rm H} / 10^{20}$ cm$^{-2}$ & $6.74$ & $6.74^\star$ && $6.74$ & $6.74^\star$ && $6.74$ & $6.74^\star$ && $6.74$ & $6.74^\star$ \\
\hline
\texttt{reflkerr} &&&&&&&& \\
$q$ & $7$ & $7.94^{+0.40}_{-0.21}$ && $7$ & $8.00^{+0.11}_{-0.06}$ && $7$ & $8.63^{+0.07}_{-0.08}$ && $7$ & $3.257^{+0.014}_{-0.017}$ \\
$i$ [deg] & $32$ & $39.3^{+0.3}_{-0.4}$ && $32$ & $40.56^{+0.15}_{-0.08}$ && $32$ & $44.84^{+0.15}_{-0.17}$ && $32$ & $34.82^{+0.30}_{-0.22}$ \\
$a_*$ & $0.85$ & $0.9092^{+0.0014}_{-0.0024}$ && $0.90$ & $0.9456^{+0.0005}_{-0.0004}$ && $0.95$ & $0.97979^{+0.00020}_{-0.00012}$ && $0.998$ & $0.9797^{+0.0007}_{-0.0008}$ \\
$A_{\rm Fe}$ & $1$ & $1.00^{+0.06}_{-0.05}$ && $1$ & $0.900^{+0.010}_{-0.012}$ && $1$ & $0.760^{+0.006}_{-0.008}$ && $1$ & $0.5000_{}^{+0.0003}$ \\
$\Gamma$ & $2$ & $1.985^{+0.010}_{-0.010}$ && $2$ & $2.0010^{+0.0010}_{-0.0043}$ && $2$ & $2.0061^{+0.0014}_{-0.0042}$ && $2$ & $1.9481^{+0.0008}_{-0.0010}$ \\
%$E_{\rm cut}$ [MeV] & $100$ & $100^{\star}$ && $100$ & $100^{\star}$ && $100$ & $100^{\star}$ && $100$ & $100^{\star}$ \\
$\mathcal{R}$ & $1$ & $0.94^{+0.06}_{-0.06}$ && 1 & $1.060^{+0.003}_{-0.035}$ && 1 & $1.114^{+0.004}_{-0.037}$ && 1 & $1.477^{+0.004}_{-0.010}$ \\
\hline
$\chi^2/\nu$ && $\quad 1031.74/992 \quad$ &&& $\quad 1130.68/992 \quad$ &&& $\quad 1484.76/992\quad$ &&& $\quad 5839.98/992 \quad$ \\
&& =1.04006 &&& =1.13980 &&& =1.49673 &&& =5.88707 \\
\hline\hline
\vspace{0.2cm}
\tablenotetext{0}{{Best-fit values for simulations~WRR1-WRR4 (returning radiation is included). The inclination angle of the disk is $i = 32^{\circ}$. The reported uncertainties correspond to 90\% confidence level for one relevant parameter. $^\star$ indicates that the parameter is frozen in the fit. The iron abundance $A_{\rm Fe}$ is allowed to vary in the range 0.5 to 10. }}
\end{tabular}}
\vspace{0.3cm}
\end{table*}

%%%%%%WRR4.1, WRR4.2, a = 0.998, q = 5, q = 3, i = 32%%%%%%
\begin{table*}
\centering
\caption{ \label{t-fit-wrr32_min}}
{\renewcommand{\arraystretch}{1.3}
\begin{tabular}{lcc|ccc}
\hline\hline
Simulation & \multicolumn{2}{c}{WRR4.1} && \multicolumn{2}{c}{WRR4.2} \\
 & Input & Fit && Input & Fit \\
\hline
\texttt{tbabs} &&&& \\
$N_{\rm H} / 10^{20}$ cm$^{-2}$ & $6.74$ & $6.74^\star$ && $6.74$ & $6.74^\star$  \\
\hline
\texttt{reflkerr} &&&& \\
$q$ & $5$ & $3.73^{+0.18}_{-0.18}$ && $3$ & $2.970^{+0.014}_{-0.016}$ \\
$i$ [deg] & $32$ & $33.3^{+1.7}_{-1.7}$ && $32$ & $32.03^{+0.17}_{-0.18}$  \\
$a_*$ & $0.998$ & $0.959^{+0.018}_{-0.024}$ && $0.998$ & $0.971^{+0.014}_{-0.014}$  \\
$A_{\rm Fe}$ & $1$ & $0.54^{+0.05}_{- (\rm P)}$ && $1$ & $0.945^{+0.020}_{-0.017}$ \\
$\Gamma$ & $2$ & $1.953^{+0.018}_{-0.018}$ && $2$ & $2.003^{+0.003}_{-0.003}$ \\
%$E_{\rm cut}$ [MeV] & $100$ & $100^{\star}$ && $100$ & $100^{\star}$ && $100$ & $100^{\star}$ && $100$ & $100^{\star}$ \\
$\mathcal{R}$ & $1$ & $1.01^{+0.11}_{-0.10}$ && $1$ & $1.089^{+0.016}_{-0.019}$ \\
\hline
$\chi^2/\nu$ && $\quad 861.52/792 \quad$ &&& $\quad 833.98/792 \quad$ \\
&& =1.0878 &&& =1.0530 \\
\hline\hline
\vspace{0.2cm}
\tablenotetext{0}{{Best-fit values for simulations~WRR4.1 and WRR4.2 (returning radiation is included). The inclination angle of the disk is $i = 32^{\circ}$. The reported uncertainties correspond to 90\% confidence level for one relevant parameter. $^\star$ indicates that the parameter is frozen in the fit. $(\rm P)$ indicates that the 90\% confidence level reaches the boundary of the parameter range (the iron abundance $A_{\rm Fe}$ is allowed to vary in the range $0.5$ to 10). }}
\end{tabular}}
\end{table*}

%%%%%%%% WRR5-WRR8 i = 57, q = 7 %%%%%%%%%%%%%%%%%%%
\begin{table*}
\centering
\caption{ \label{t-fit-b}}
{\renewcommand{\arraystretch}{1.3}
\begin{tabular}{lcc|ccc|ccc|ccc}
\hline\hline
Simulation & \multicolumn{2}{c}{WRR5} && \multicolumn{2}{c}{WRR6} && \multicolumn{2}{c}{WRR7} && \multicolumn{2}{c}{WRR8} \\
 & Input & Fit && Input & Fit && Input & Fit && Input & Fit \\
\hline
\texttt{tbabs} &&&&&&&&&& \\
$N_{\rm H} / 10^{20}$ cm$^{-2}$ & $6.74$ & $6.74^\star$ && $6.74$ & $6.74^\star$ && $6.74$ & $6.74^\star$ && $6.74$ & $6.74^\star$ \\
\hline
\texttt{reflkerr} &&&&&&&& \\
$q$ & $7$ & $7.8^{+0.8}_{-0.8}$ && $7$ & $7.60^{+0.20}_{-0.15}$ && $7$ & $6.50^{+0.10}_{-0.06}$ && $7$ & $4.88^{+0.12}_{-0.07}$ \\
$i$ [deg] & $57$ & $57.5^{+0.4}_{-0.6}$ && $57$ & $58.00^{+0.13}_{-0.10}$ && $57$ & $55.71^{+0.10}_{-0.14}$ && $57$ & $56.8^{+0.4}_{-0.4}$ \\
$a_*$ & $0.85$ & $0.858^{+0.009}_{-0.015}$ && $0.90$ & $0.9089^{+0.0006}_{-0.0010}$ && $0.95$ & $0.9457^{+0.0005}_{-0.0004}$ && $0.998$ & $0.9932^{+0.0017}_{-0.0016}$ \\
$A_{\rm Fe}$ & $1$ & $1.09^{+0.12}_{-0.08}$ && $1$ & $0.971^{+0.015}_{-0.030}$ && $1$ & $0.942^{+0.013}_{-0.015}$ && $1$ & $0.500_{}^{+0.003}$ \\
$\Gamma$ & $2$ & $2.000^{+0.005}_{-0.004}$ && $2$ & $2.0025^{+0.0014}_{-0.0018}$ && $2$ & $2.005^{+0.0005}_{-0.0005}$ && $2$ & $2.000^{+0.004}_{-0.005}$ \\
%$E_{\rm cut}$ [MeV] & $100$ & $100^{\star}$ && $100$ & $100^{\star}$ && $100$ & $100^{\star}$ && $100$ & $100^{\star}$ \\
$\mathcal{R}$ & $1$ & $1.02^{+0.03}_{-0.04}$ && 1 & $1.051^{+0.011}_{-0.013}$ && 1 & $1.077^{+0.004}_{-0.003}$ && 1 & $1.113^{+0.021}_{-0.022}$ \\
\hline
$\chi^2/\nu$ && $\quad 984.55/992 \quad$ &&& $\quad 921.92/992 \quad$ &&& $\quad 983.94/992 \quad$ &&& $\quad 1233.18/992 \quad$ \\
&& =0.99249 &&& =0.92936 &&& =0.99188 &&& =1.24313 \\
\hline\hline
\vspace{0.2cm}
\tablenotetext{0}{{Best-fit values for simulations~WRR5-WRR8 (returning radiation is included). The inclination angle of the disk is $i = 57^{\circ}$. The reported uncertainties correspond to 90\% confidence level for one relevant parameter. $^\star$ indicates that the parameter is frozen in the fit. The iron abundance $A_{\rm Fe}$ is allowed to vary in the range 0.5 to 10. }}
\end{tabular}
}
\end{table*}

%%%%WRR8.1, WRR8.2, a = 0.998, q = 5, q = 3, i = 57%%%%%%%%%%%%
\begin{table*}
\centering
\caption{ \label{t-fit-b_min}}
{\renewcommand{\arraystretch}{1.3}
\begin{tabular}{lcc|ccc}
\hline\hline
Simulation & \multicolumn{2}{c}{WRR8.1} && \multicolumn{2}{c}{WRR8.2} \\
 & Input & Fit && Input & Fit \\
\hline
\texttt{tbabs} &&&& \\
$N_{\rm H} / 10^{20}$ cm$^{-2}$ & $6.74$ & $6.74^\star$ && $6.74$ & $6.74^\star$  \\
\hline
\texttt{reflkerr} &&&& \\
$q$ & $5$ & $4.43^{+0.08}_{-0.07}$ && $3$ & $2.98^{+0.06}_{-0.03}$ \\
$i$ [deg] & $57$ & $56.6^{+0.4}_{-0.3}$ && $57$ & $56.87^{+0.14}_{-0.10}$  \\
$a_*$ & $0.998$ & $0.9956^{+ (\rm P)}_{-0.0005}$ && $0.998$ & $0.992^{+ (\rm P)}_{-0.005}$  \\
$A_{\rm Fe}$ & $1$ & $0.793^{+0.019}_{-0.015}$ && $1$ & $0.956^{+0.028}_{-0.021}$ \\
$\Gamma$ & $2$ & $2.0047^{+0.0027}_{-0.0010}$ && $2$ & $2.0018^{+0.0019}_{-0.0022}$ \\
%$E_{\rm cut}$ [MeV] & $100$ & $100^{\star}$ && $100$ & $100^{\star}$ && $100$ & $100^{\star}$ && $100$ & $100^{\star}$ \\
$\mathcal{R}$ & $1$ & $1.198^{+0.024}_{-0.006}$ && $1$ & $1.065^{+0.026}_{-0.014}$ \\
\hline
$\chi^2/\nu$ && $\quad 789.00/792 \quad$ &&& $\quad 830.28/792 \quad$ \\
&& =0.99621 &&& =1.0483 \\
\hline\hline
\vspace{0.2cm}
\tablenotetext{0}{{Best-fit values for simulations~WRR8.1 and WRR8.2 (returning radiation is included). The inclination angle of the disk is $i = 57^{\circ}$. The reported uncertainties correspond to 90\% confidence level for one relevant parameter. $^\star$ indicates that the parameter is frozen in the fit. $(\rm P)$ indicates that the 90\% confidence level reaches the boundary of the parameter range (the spin parameter $a_*$ is allowed to vary in the range $-0.998$ to 0.998). }}
\end{tabular}}
\end{table*}

%%%%%%%% WRR9-WRR11, i = 32, broken power-law %%%%%%%%%%%%%%%%%%%
\begin{table*}
\centering
\caption{ \label{t-fit-rbr-i32}}
{\renewcommand{\arraystretch}{1.3}
\begin{tabular}{lcc|ccc|ccc}
\hline\hline
Simulation & \multicolumn{2}{c}{WRR9} && \multicolumn{2}{c}{WRR10} && \multicolumn{2}{c}{WRR11} \\
 & Input & Fit && Input & Fit && Input & Fit  \\
\hline
\texttt{tbabs} &&&&&&&& \\
$N_{\rm H} / 10^{20}$ cm$^{-2}$ & $6.74$ & $6.74^\star$ && $6.74$ & $6.74^\star$ && $6.74$ & $6.74^\star$  \\
\hline
\texttt{reflkerr} &&&&&&&& \\
$q_{\rm in}$ & $7$ & $7.17^{+0.27}_{-0.17}$ && $7$ & $6.90^{+0.15}_{-0.21}$ && $7$ & $7.15^{+0.07}_{-0.12}$ \\
$q_{\rm out}$ & $3$ & $2.970^{+0.025}_{-0.022}$ && $3$ & $2.851^{+0.019}_{-0.029}$ && $3$ & $3.085^{+0.018}_{-0.025}$ \\
$r_{\rm br}$ [$R_{\rm g}$] & $4$ & $3.85^{+0.08}_{-0.11}$ && $4$ & $3.54^{+0.11}_{-0.08}$ && $3$ & $3.32^{+0.04}_{-0.05}$ \\
$i$ [deg] & $32$ & $31.92^{+0.18}_{-0.21}$ && $32$ & $31.45^{+0.17}_{-0.25}$ && $32$ & $31.95^{+0.18}_{-0.12}$  \\
$a_*$ & $0.9$ & $0.9098^{+0.0021}_{-0.0025}$ && $0.998$ & $0.9708^{+0.0005}_{-0.0010}$ && $0.998$ & $0.9796^{+0.0006}_{-0.0004}$  \\
$A_{\rm Fe}$ & $1$ & $0.978^{+0.014}_{-0.013}$ && $1$ & $0.542^{+0.010}_{-0.020}$ && $1$ & $0.878^{+0.013}_{-0.010}$  \\
$\Gamma$ & $2$ & $2.002^{+0.004}_{-0.005}$ && $2$ & $2.046^{+0.007}_{-0.013}$ && $2$ & $2.151^{+0.019}_{-0.023}$ \\
%$E_{\rm cut}$ [MeV] & $100$ & $100^{\star}$ && $100$ & $100^{\star}$ && $100$ & $100^{\star}$ && $100$ & $100^{\star}$ \\
$\mathcal{R}$ & $1$ & $1.03^{+0.03}_{-0.03}$ && 1 & $1.87^{+0.05}_{-0.09}$ && 1 & $2.56^{+0.14}_{-0.204}$ \\
\hline
$\chi^2/\nu$ && $\quad 870.31/790 \quad$ &&& $\quad 1161.76/790 \quad$ &&& $\quad 911.29/790 \quad$  \\
&& =1.1017 &&& =1.47058 &&& =1.1535 \\
\hline\hline
\vspace{0.2cm}
\tablenotetext{0}{{Best-fit values for simulations~WRR9-WRR11 (returning radiation is included). The inclination angle of the disk is $i = 32^{\circ}$. The reported uncertainties correspond to 90\% confidence level for one relevant parameter. $^\star$ indicates that the parameter is frozen in the fit. }}
\end{tabular}
}
\end{table*}

%%%%%%%% WRR9-WRR11, i = 57, broken power-law %%%%%%%%%%%%%%%%%%%
\begin{table*}
\centering
\caption{ \label{t-fit-rbr-i57}}
{\renewcommand{\arraystretch}{1.3}
\begin{tabular}{lcc|ccc|ccc}
\hline\hline
Simulation & \multicolumn{2}{c}{WRR12} && \multicolumn{2}{c}{WRR13} && \multicolumn{2}{c}{WRR14} \\
 & Input & Fit && Input & Fit && Input & Fit  \\
\hline
\texttt{tbabs} &&&&&&&& \\
$N_{\rm H} / 10^{20}$ cm$^{-2}$ & $6.74$ & $6.74^\star$ && $6.74$ & $6.74^\star$ && $6.74$ & $6.74^\star$  \\
\hline
\texttt{reflkerr} &&&&&&&& \\
$q_{\rm in}$ & $7$ & $7.5^{+0.6}_{-0.7}$ && $7$ & $5.2^{+0.3}_{-0.3}$ && $7$ & $6.3^{+0.3}_{-0.7}$ \\
$q_{\rm out}$ & $3$ & $2.83^{+0.50}_{-0.12}$ && $3$ & $3.88^{+0.29}_{-0.74}$ && $3$ & $3.00^{+0.13}_{-0.10}$ \\
$r_{\rm br}$ [$R_{\rm g}$] & $4$ & $3.81^{+0.26}_{-0.19}$ && $4$ & $2.8^{+0.9}_{-0.4}$ && $3$ & $2.72^{+0.09}_{-0.09}$ \\
$i$ [deg] & $57$ & $56.89^{+0.15}_{-0.15}$ && $57$ & $56.8^{+0.4}_{-0.3}$ && $57$ & $56.79^{+0.25}_{-0.21}$  \\
$a_*$ & $0.9$ & $0.9092^{+0.0017}_{-0.0037}$ && $0.998$ & $0.9848^{+0.0050}_{-0.0026}$ && $0.998$ & $0.9951^{+0.0024}_{-0.0034}$  \\
$A_{\rm Fe}$ & $1$ & $0.964^{+0.028}_{-0.023}$ && $1$ & $0.511^{+0.013}_{-\rm (P)}$ && $1$ & $0.692^{+0.020}_{-0.060}$  \\
$\Gamma$ & $2$ & $1.9994^{+0.0037}_{-0.0023}$ && $2$ & $1.968^{+0.006}_{-0.004}$ && $2$ & $1.996^{+0.009}_{-0.014}$ \\
%$E_{\rm cut}$ [MeV] & $100$ & $100^{\star}$ && $100$ & $100^{\star}$ && $100$ & $100^{\star}$ && $100$ & $100^{\star}$ \\
$\mathcal{R}$ & $1$ & $1.011^{+0.013}_{-0.014}$ && 1 & $1.058^{+0.033}_{-0.026}$ && 1 & $1.21^{+0.04}_{-0.09}$ \\
\hline
$\chi^2/\nu$ && $\quad 745.61/790 \quad$ &&& $\quad 855.44/790 \quad$ &&& $\quad 813.57/790 \quad$  \\
&& =0.94381 &&& =1.0828 &&& =1.0298 \\
\hline\hline
\vspace{0.2cm}
\tablenotetext{0}{{Best-fit values for simulations~WRR12-WRR14 (returning radiation is included). The inclination angle of the disk is $i = 57^{\circ}$. The reported uncertainties correspond to 90\% confidence level for one relevant parameter. $^\star$ indicates that the parameter is frozen in the fit. The iron abundance $A_{\rm Fe}$ is allowed to vary in the range 0.5 to 10. }}
\end{tabular}
}
\end{table*}

%%%%%%%%%%% NuSTAR cases %%%%%%%%%

\begin{table*}
\centering
\caption{ \label{t-fit-ns}}
{\renewcommand{\arraystretch}{1.3}
\begin{tabular}{lcc|ccc|ccc|ccc}
\hline\hline
Simulation & \multicolumn{2}{c}{NRR4.1} && \multicolumn{2}{c}{WRR4.1} && \multicolumn{2}{c}{NRR4.2} && \multicolumn{2}{c}{WRR4.2} \\
 & Input & Fit && Input & Fit && Input & Fit && Input & Fit \\
\hline
\texttt{tbabs} &&&& \\
$N_{\rm H} / 10^{20}$ cm$^{-2}$ & $6.74$ & $6.74^\star$ && $6.74$ & $6.74^\star$ && $6.74$ & $6.74^\star$ && $6.74$ & $6.74^\star$ \\
\hline
\texttt{reflkerr} &&&& \\
$q$ & $5$ & $5.36^{+0.22}_{-0.19}$ && $5$ & $9.56^{+0.04}_{-0.04}$ && $3$ & $3.04^{+0.03}_{-0.03}$ && $3$ & $3.201^{+0.023}_{-0.022}$ \\
$i$ [deg] & $32$ & $35.4^{+1.9}_{-1.8}$ && $32$ & $61.48^{+0.05}_{-0.10}$ && $32$ & $32.00^{+0.04}_{-0.04}$ && $32$ & $32.9^{+0.3}_{-0.4}$ \\
$a_*$ & $0.998$ & $0.9980^{}_{-0.0006}$ && $0.998$ & $0.99563600^{+0.000031}_{-0.000009}$ && $0.998$ & $0.978^{+0.018}_{-0.021}$ && $0.998$ & $0.929^{+0.005}_{-0.004}$ \\
$A_{\rm Fe}$ & $1$ & $1.16^{+0.05}_{-0.06}$ && $1$ & $3.155^{+0.017}_{-0.018}$ && $1$ & $1.08^{+0.05}_{-0.05}$ && $1$ & $1.239^{+0.019}_{-0.015}$ \\
$\Gamma$ & $2$ & $1.994^{+0.003}_{-0.003}$ && $2$ & $1.7980^{+0.0007}_{-0.0007}$ && $2$ & $1.985^{+0.005}_{-0.006}$ && $2$ & $1.959^{+0.003}_{-0.005}$ \\
%$E_{\rm cut}$ [MeV] & $100$ & $100^{\star}$ && $100$ & $100^{\star}$ && $100$ & $100^{\star}$ && $100$ & $100^{\star}$ \\
$\mathcal{R}$ & $1$ & $1.033^{+0.036}_{-0.017}$ && $1$ & $1.585^{+0.005}_{-0.006}$ && $1$ & $0.95^{+0.03}_{-0.03}$ && $1$ & $1.080^{+0.006}_{-0.016}$ \\
\hline
$\chi^2/\nu$ && $\quad 1351.19/1343 \quad$ &&& $\quad 3061.54/1340 \quad$ &&& $\quad 1372.68/1339 \quad$ &&& $\quad 1643.88/1339 \quad$ \\
&& =1.00610 &&& =2.28473 &&& =1.02516 &&& =1.22769 \\
\hline\hline
\vspace{0.2cm}
\tablenotetext{0}{{Best-fit values for the simulations with \textsl{NuSTAR} with the input parameters of NRR4.1, WRR4.1, NRR4.2, and WRR4.2. The reported uncertainties correspond to 90\% confidence level for one relevant parameter. $^\star$ indicates that the parameter is frozen in the fit. The spin parameter $a_*$ is allowed to vary in the range $-0.998$ to 0.998. Note that we only analyze the energy range 3-30~keV (see text for more details).}}
\end{tabular}}
\end{table*}

%%%%WRR41, a = 0.998, q = 5, i = 32 NuSTAR broken power-law%%%%%%%%%%%%
\begin{table*}
\centering
\caption{ \label{ns-bpl}}
{\renewcommand{\arraystretch}{1.3}
\begin{tabular}{lcc}
\hline\hline
Simulation & \multicolumn{2}{c}{WRR4.1}  \\
 & Input & Fit  \\
\hline
\texttt{tbabs} && \\
$N_{\rm H} / 10^{20}$ cm$^{-2}$ & $6.74$ & $6.74^\star$   \\
\hline
\texttt{reflkerr} && \\
$q$ & $5$ & $-$  \\
$q_{\rm in}$ & $-$ & $10.00^{}_{-0.01}$ \\
$q_{\rm out}$ & $-$ & $5.05^{+0.02}_{-0.02}$ \\
$r_{\rm br}$ & $-$ & $2.494^{+0.004}_{-0.005}$ \\
$i$ [deg] & $32$ & $41.50^{+0.06}_{-0.08}$   \\
$a_*$ & $0.998$ & $0.97971^{+0.00004}_{-0.00004}$   \\
$A_{\rm Fe}$ & $1$ & $9.19^{+0.03}_{-0.05}$ \\
$\Gamma$ & $2$ & $1.633^{+0.003}_{-0.001}$  \\
%$E_{\rm cut}$ [MeV] & $100$ & $100^{\star}$ && $100$ & $100^{\star}$ && $100$ & $100^{\star}$ && $100$ & $100^{\star}$ \\
$\mathcal{R}$ & $1$ & $1.250^{+0.004}_{-0.003}$ \\
\hline
$\chi^2/\nu$ && $\quad 1880.0/1338 \quad$  \\
&& =1.4051 \\
\hline\hline
\vspace{0.2cm}
\tablenotetext{0}{{Best-fit values for simulations~WRR4.1 (returning radiation is included) simulated with \textsl{NuSTAR} and fitted with the broken power-law. The inclination angle of the disk is $i = 32^{\circ}$. The reported uncertainties correspond to 90\% confidence level for one relevant parameter. $^\star$ indicates that the parameter is frozen in the fit. The iron abundance $A_{\rm Fe}$ is allowed to vary in the range 0.5 to 10. }}
\end{tabular}}
\end{table*}


\begin{thebibliography}{99}

\bibitem[Abdikamalov et al.(2020)]{2020arXiv200309663A} Abdikamalov, A.~B., Ayzenberg, D., Bambi, C., et al.\ 2020, \apj, 899, 80

\bibitem[Agol \& Krolik(2000)]{AK2000} Agol, E., \& Krolik, J.~H.\ 2000, \apj, 528, 161

\bibitem[Arnaud(1996)]{xspec} Arnaud, K.~A.\ 1996, Astronomical Data Analysis Software and Systems V, 17

\bibitem[Bambi(2017)]{2017RvMP...89b5001B} Bambi, C.\ 2017, Reviews of Modern Physics, 89, 025001

\bibitem[Bambi et al.(2017)]{2017ApJ...842...76B} Bambi, C., C{\'a}rdenas-Avenda{\~n}o, A., Dauser, T., et al.\ 2017, \apj, 842, 76

\bibitem[Brenneman \& Reynolds(2006)]{2006ApJ...652.1028B} Brenneman, L.~W., \& Reynolds, C.~S.\ 2006, \apj, 652, 1028

\bibitem[Brenneman et al.(2013)]{2013MNRAS.429.2662B} Brenneman, L.~W., Risaliti, G., Elvis, M., et al.\ 2013, \mnras, 429, 2662

\bibitem[Cao et al.(2018)]{2018PhRvL.120e1101C} Cao, Z., Nampalliwar, S., Bambi, C., et al.\ 2018, \prl, 120, 051101

\bibitem[Cardenas-Avendano et al.(2020)]{2020arXiv200506719C} C{\'a}rdenas-Avenda{\~n}o, A., Zhou, M., \& Bambi, C.\ 2020, \prd, 101, 123014

\bibitem[Cunningham(1976)]{CT} Cunningham, C.\ 1976, \apj, 208, 534

\bibitem[Dabrowski et al.(1997)]{Dab1997} Dabrowski, Y., Fabian, A.~C., Iwasawa, K., et al.\ 1997, \mnras, 288, L11

\bibitem[De Rosa et al.(2019)]{2019SCPMA..6229504D} De Rosa, A., Uttley, P., Gou, L., et al.\ 2019, Science China Physics, Mechanics, and Astronomy, 62, 29504

\bibitem[Duro et al.(2016)]{2016A&A...589A..14D} Duro, R., Dauser, T., Grinberg, V., et al.\ 2016, \aap, 589, A14


\bibitem[El-Batal et al.(2016)]{2016ApJ...826L..12E} El-Batal, A.~M., Miller, J.~M., Reynolds, M.~T., et al.\ 2016, \apjl, 826, L12

\bibitem[Fabian et al.(2012)]{2012MNRAS.419..116F} Fabian, A.~C., Zoghbi, A., Wilkins, D., et al.\ 2012, \mnras, 419, 116

\bibitem[Frolov et al.(2014)]{2014JCAP...07..059F} Frolov, V.~P., Shoom, A.~A., \& Tzounis, C.\ 2014, \jcap, 2014, 059

\bibitem[Garc{\'\i}a \& Kallman(2010)]{2010ApJ...718..695G} Garc{\'\i}a, J., \& Kallman, T.~R.\ 2010, \apj, 718, 695

\bibitem[Garc{\'\i}a et al.(2013)]{2013ApJ...768..146G} Garc{\'\i}a, J., Dauser, T., Reynolds, C.~S., et al.\ 2013, \apj, 768, 146

\bibitem[Garc{\'\i}a et al.(2018)]{2018ApJ...864...25G} Garc{\'\i}a, J.~A., Steiner, J.~F., Grinberg, V., et al.\ 2018, \apj, 864, 25

\bibitem[Garc{\'\i}a et al.(2018)]{2018ASPC..515..282G} Garc{\'\i}a, J.~A., Kallman, T.~R., Bautista, M., et al.\ 2018, Workshop on Astrophysical Opacities, 282

\bibitem[Gendreau et al.(2017)]{nicer} Gendreau, K., Arzoumanian, Z., \& NICER Team\ 2017, American Astronomical Society Meeting Abstracts \#229 229, 309.03

\bibitem[George \& Fabian(1991)]{1991MNRAS.249..352G} George, I.~M., \& Fabian, A.~C.\ 1991, \mnras, 249, 352

\bibitem[Gorecki \& Wilczewski(1984)]{1984AcA....34..141G} Gorecki, A., \& Wilczewski, W.\ 1984, \actaa, 34, 141

\bibitem[Grevesse \& Sauval(1998)]{1998SSRv...85..161G} Grevesse, N., \& Sauval, A.~J.\ 1998, \ssr, 85, 161

\bibitem[Johannsen \& Psaltis(2013)]{2013ApJ...773...57J} Johannsen, T. \& Psaltis, D.\ 2013, \apj, 773, 57

\bibitem[Kammoun et al.(2018)]{2018A&A...614A..44K} Kammoun, E.~S., Nardini, E., \& Risaliti, G.\ 2018, \aap, 614, A44

\bibitem[Keck et al.(2015)]{2015ApJ...806..149K} Keck, M.~L., Brenneman, L.~W., Ballantyne, D.~R., et al.\ 2015, \apj, 806, 149

\bibitem[Li et al.(2005)]{2005ApJS..157..335L} Li, L.-X., Zimmerman, E.~R., Narayan, R., et al.\ 2005, \apjs, 157, 335

\bibitem[Lu \& Torres(2003)]{2003IJMPD..12...63L} Lu, Y. \& Torres, D.~F.\ 2003, International Journal of Modern Physics D, 12, 63

\bibitem[Magdziarz \& Zdziarski(1995)]{1995MNRAS.273..837M} Magdziarz, P., \& Zdziarski, A.~A.\ 1995, \mnras, 273, 837


\bibitem[Marinucci et al.(2014)]{2014ApJ...787...83M} Marinucci, A., Matt, G., Miniutti, G., et al.\ 2014, \apj, 787, 83

\bibitem[Miller et al.(2013)]{2013ApJ...775L..45M} Miller, J.~M., Parker, M.~L., Fuerst, F., et al.\ 2013, \apjl, 775, L45

\bibitem[Miniutti et al.(2003)]{2003MNRAS.344L..22M} Miniutti, G., Fabian, A.~C., Goyder, R., et al.\ 2003, \mnras, 344, L22

\bibitem[Morrison \& McCammon(1983)]{1983ApJ...270..119M} Morrison, R., \& McCammon, D.\ 1983, \apj, 270, 119


\bibitem[Nied{\'z}wiecki(2005)]{Nied2005} Nied{\'z}wiecki, A.\ 2005, \mnras, 356, 913 
 
\bibitem[Nied{\'z}wiecki \& {\.Z}ycki(2008)]{Nied2008} Nied{\'z}wiecki, A., \& {\.Z}ycki, P.~T.\ 2008, \mnras, 386, 759

\bibitem[Nied{\'z}wiecki et al.(2016)]{2016ApJ...821L...1N} Nied{\'z}wiecki, A., Zdziarski, A.~A., \& Szanecki, M.\ 2016, \apjl, 821, L1

\bibitem[Nied{\'z}wiecki \& Zdziarski(2018)]{2018MNRAS.477.4269N} Nied{\'z}wiecki, A., \& Zdziarski, A.~A.\ 2018, \mnras, 477, 4269

\bibitem[Nied{\'z}wiecki et al.(2019)]{reflkerr} Nied{\'z}wiecki, A., Szanecki, M., \& Zdziarski, A.~A.\ 2019, \mnras, 485, 2942

\bibitem[Noble et al.(2010)]{2010ApJ...711..959N} Noble, S.~C., Krolik, J.~H., \& Hawley, J.~F.\ 2010, \apj, 711, 959

\bibitem[Penna et al.(2012)]{2012MNRAS.420..684P} Penna, R.~F., Sadowski, A., \& McKinney, J.~C.\ 2012, \mnras, 420, 684

\bibitem[Pozdnyakov et al.(1983)]{1983ASPRv...2..189P} Pozdnyakov, L.~A., Sobol, I.~M., \& Syunyaev, R.~A.\ 1983, \apspr, 2, 189

\bibitem[Reynolds \& Begelman(1997)]{1997ApJ...488..109R} Reynolds, C.~S. \& Begelman, M.~C.\ 1997, \apj, 488, 109

\bibitem[Reynolds \& Fabian(2008)]{2008ApJ...675.1048R} Reynolds, C.~S., \& Fabian, A.~C.\ 2008, \apj, 675, 1048

\bibitem[Reynolds et al.(2012)]{2012ApJ...755...88R} Reynolds, C.~S., Brenneman, L.~W., Lohfink, A.~M., et al.\ 2012, \apj, 755, 88

\bibitem[Riaz et al.(2020a)]{2020MNRAS.491..417R} Riaz, S., Ayzenberg, D., Bambi, C., et al.\ 2020a, \mnras, 491, 417

\bibitem[Riaz et al.(2020b)]{2019arXiv191106605R} Riaz, S., Ayzenberg, D., Bambi, C., et al.\ 2020b, \apj, 895, 61

\bibitem[Risaliti et al.(2013)]{2013Natur.494..449R} Risaliti, G., Harrison, F.~A., Madsen, K.~K., et al.\ 2013, \nat, 494, 449

\bibitem[Ross \& Fabian(2005)]{2005MNRAS.358..211R} Ross, R.~R., \& Fabian, A.~C.\ 2005, \mnras, 358, 211

\bibitem[Ross et al.(2002)]{2002MNRAS.336..315R} Ross, R.~R., Fabian, A.~C., \& Ballantyne, D.~R.\ 2002, \mnras, 336, 315

\bibitem[Schee \& Stuchl{\'\i}k(2009)]{2009GReGr..41.1795S} Schee, J. \& Stuchl{\'\i}k, Z.\ 2009, General Relativity and Gravitation, 41, 1795

\bibitem[Schnittman \& Krolik(2009)]{2009ApJ...701.1175S} Schnittman, J.~D., \& Krolik, J.~H.\ 2009, \apj, 701, 1175

\bibitem[Suebsuwong et al.(2006)]{sue2006} Suebsuwong, T., Malzac, J., Jourdain, E., et al.\ 2006, \aap, 453, 773

\bibitem[Sunyaev \& Truemper(1979)]{1979Natur.279..506S} Sunyaev, R.~A., \& Truemper, J.\ 1979, \nat, 279, 506

\bibitem[Szanecki et al.(2020)]{2020arXiv200615016S} Szanecki, M., Nied{\'z}wiecki, A., Done, C., et al.\ 2020, \aap, 641, A89

\bibitem[Taylor \& Reynolds(2018)]{2018ApJ...855..120T} Taylor, C., \& Reynolds, C.~S.\ 2018, \apj, 855, 120

\bibitem[Tomsick et al.(2014)]{2014ApJ...780...78T} Tomsick, J.~A., Nowak, M.~A., Parker, M., et al.\ 2014, \apj, 780, 78

\bibitem[Tripathi et al.(2019)]{2019ApJ...875...56T} Tripathi, A., Nampalliwar, S., Abdikamalov, A.~B., et al.\ 2019, \apj, 875, 56

\bibitem[Walton et al.(2013)]{2013MNRAS.428.2901W} Walton, D.~J., Nardini, E., Fabian, A.~C., et al.\ 2013, \mnras, 428, 2901

\bibitem[Wilkins \& Fabian(2012)]{WF2012} Wilkins, D.~R., \& Fabian, A.~C.\ 2012, \mnras, 424, 1284

\bibitem[Wilkins \& Gallo(2015)]{2015MNRAS.449..129W} Wilkins, D.~R. \& Gallo, L.~C.\ 2015, \mnras, 449, 129

\bibitem[Wilkins et al.(2020a)]{2020MNRAS.493.5532W} Wilkins, D.~R., Reynolds, C.~S., \& Fabian, A.~C.\ 2020, \mnras, 493, 5532

\bibitem[Wilkins et al.(2020b)]{2020MNRAS.498.3302W} Wilkins, D.~R., Garc{\'\i}a, J.~A., Dauser, T., et al.\ 2020, \mnras, 498, 3302

\bibitem[Wilms et al.(2000)]{wilms} Wilms, J., Allen, A., \& McCray, R.\ 2000, \apj, 542, 914

\bibitem[Xu et al.(2018)]{2018ApJ...865..134X} Xu, Y., Nampalliwar, S., Abdikamalov, A.~B., et al.\ 2018, \apj, 865, 134

\bibitem[Zhang et al.(2019)]{2019ApJ...884..147Z} Zhang, Y., Abdikamalov, A.~B., Ayzenberg, D., et al.\ 2019, \apj, 884, 147

\bibitem[Zhou et al.(2020)]{2020PhRvD.101d3010Z} Zhou, M., Ayzenberg, D., Bambi, C., et al.\ 2020, \prd, 101, 043010



\end{thebibliography}
\end{document}